\documentclass{article}

\usepackage{comment}
\usepackage{ulem}
\usepackage{cancel}
\usepackage{subcaption} % Pacote para subfiguras modernas

\usepackage{float}
\usepackage{graphicx} 
\usepackage{cite}
\usepackage{booktabs}
\usepackage{amsmath}
\usepackage{amssymb}
\usepackage{graphicx}
\usepackage{esint}
\usepackage{hyperref}
\usepackage{multirow}

\usepackage{color}

\usepackage{tikz}
\usepackage{tikz-feynman}
\tikzfeynmanset{compat=1.1.0}

\title{
A microscopic stochastic field theory for nonlinear pomeron--odderon evolution based on a Doi--Peliti framework}

\author{Anderson Kendi Kohara}

\date{    $^1$CPHT, CNRS, École Polytechnique, Institut Polytechnique de Paris, 91120 Palaiseau, France \footnote{email:anderson.kendi@gmail.com}
\\%
}

\begin{document}

\maketitle

\begin{abstract}

We construct a reaction-based microscopic realisation of nonlinear
pomeron--odderon evolution in rapidity and impact-parameter space.
Using the Doi--Peliti formalism, we derive the master equations,
their operator representation, the exact hierarchy of factorial
moments, and the coherent-state functional integral. The construction
admits normalised complex configuration weights while retaining
the algebraic conservation law of the gain--loss generator.
Pomeron branching and merging yield the complex logistic equation \cite{Kohara:2026}
under mean-field factorisation. Pomeron-catalysed odderon branching
and ternary odderon fusion provide an odd sector whose evolution
is driven by the pomeron background. Without transverse diffusion,
the resulting equations admit an analytical solution in quadrature,
allowing us to examine amplification, phase evolution, and nonlinear
regulation. A specified matching between the first moments and
the even and odd scattering profiles connects this evolution to
elastic-scattering observables. Illustrative initial conditions
produce a nonmonotonic forward ratio and a splitting between the
$pp$ and $p\bar p$ channels, while the interference near a diffractive
minimum remains sensitive to the relative phases and transverse
profiles. These examples illustrate the effective dynamics without
constituting a fit to data or a general proof of scattering unitarity.

\end{abstract}

\section{Introduction}

High-energy hadronic scattering remains one of the most challenging regimes of quantum chromodynamics (QCD), where non-perturbative dynamics and unitarity constraints interplay in a nontrivial way. In the soft regime, the non-perturbative nature of hadronic elastic scattering makes a direct calculation of scattering amplitudes from QCD degrees of freedom particularly difficult. A phenomenological description is therefore usually required. A large number of models \cite{Islam:2009,kohara:2014,Kohara:2019qoq,Csorgo:2013usx,BSW:1979,Selyugin:2023txo,Donnachie:2013xia} describe observables over an energy range extending from a few GeV to tens of TeV. Despite this vast literature \cite{Collins:1977,Abarbanel:1975}, a unified description connecting soft elastic scattering to the underlying QCD dynamics remains an open problem.

Regge theory \cite{Regge:1959,Chew:1961,Gribov:1961} and its field-theoretic formulation, Reggeon Field Theory (RFT) \cite{Gribov:1967,Abarbanel:1975}, provide an effective description in terms of exchanged collective degrees of freedom. These exchanges can be classified according to their charge-conjugation quantum numbers. The  eigenvalue of a physical state under charge conjugation operation is +1 for $C$-even and -1 for $C$-odd. Secondary Reggeons contribute significantly at lower energies, while the pomeron describes the dominant high-energy, $C$-even contribution. The odderon describes a $C$-odd contribution that changes sign between proton--proton and proton--antiproton amplitudes. Its possible role in elastic scattering has received renewed attention following measurements of the forward $\rho$ parameter and comparisons between the two scattering channels \cite{Antchev:2019,Abazov:2021}.

Regge theory is formulated through the analytic continuation of crossed-channel partial waves to complex angular momentum. The singularities in this plane determine the high-energy behaviour of the amplitude. In phenomenological applications, a limited set of trajectories and residues is retained, and their parameters must be determined from data. RFT extends this description by organizing Reggeon propagation and interactions in terms of Feynman-like diagrams \cite{Gribov:1967}. Connections with QCD have been established in perturbative high-energy domain, including coupled pomeron--odderon evolution in the linear diluted regime \cite{Kwiecinski:1980, Bartels:1980, Ewerz:2003xi} and also in the dense regime accounting for gluons saturation \cite{Kovchegov:2004,Hatta:2005}. However, a quantitative description of soft hadronic elastic amplitudes directly from QCD remains difficult.

We have previously proposed an approach to hadronic elastic scattering for the pomeron sector that transforms its energy dependence into an initial value problem \cite{Kakkad:2022}. More recently, we showed how a linear evolution equation reproduces the energy dependence of a single pomeron-like contribution and introduced a nonlinear term to control its growth \cite{Kohara:2026}. The resulting complex logistic equation admits an analytical solution in closed form. Once an initial profile is specified, its energy evolution is governed by the effective pomeron intercept parameter $\epsilon_{\mathcal P}$ and the nonlinear coupling $\lambda$. In RFT language, the intercept parameter corresponds, up to the sign convention, to the quadratic mass coefficient. This formulation separates the choice of the initial amplitude profile from the law governing its evolution.

In the present work, we address three main questions:
\begin{itemize}
\item Can we construct a microscopic model giving rise to this evolution equation?
\item How does the nonlinear term emerge from the microscopic dynamics?
\item How can we minimally include an odderon field, and what are the consequences for scattering observables?
\end{itemize}

To address these questions, we revisit the pomeron--odderon system from a microscopic perspective by constructing a local reaction--diffusion framework. The relation between RFT and stochastic processes has a long history \cite{Grassberger:1978,Cardy:1980}. Reaction--diffusion representations have been studied in zero transverse dimensions \cite{Bondarenko:2007} and used to calculate RFT amplitudes with applications to elastic scattering \cite{Kolevatov:2011}. Coupled pomeron--odderon field theories have also been investigated \cite{Bartels:2017,Braun:2021}, while the difficulties of relating their field variables to positive probabilities have recently been examined explicitly \cite{Braun:2025}.

Our aim is to construct a microscopic realization of the complex logistic evolution together with a dynamical odd sector. We use the Doi--Peliti formalism \cite{Doi:1976,Peliti:1985}, which provides a systematic mapping from master equations to operator and field-theoretic representations; detailed accounts can be found in Refs.~\cite{Tauber:2005,Weber:2017}. The local network combines pomeron branching and merging, $P\to 2P$ and $2P\to P$, with pomeron-catalysed odderon branching, $P+O\to P+3O$, and ternary odderon fusion, $3O\to O$. Diffusion accounts for propagation in the transverse impact-parameter plane. The catalytic channel leaves the pomeron multiplicity unchanged while increasing the odderon multiplicity by two, and therefore amplifies an existing odd component rather than providing a source in its absence.

A key feature of this construction is that the Hamiltonian is obtained from specified gain--loss equations rather than postulated directly. Each reaction fixes both the interaction vertex and its associated loss term, allowing us to control the combinatorial structure and conservation of the total weight. For positive initial weights and non-negative rates, these equations define a stochastic process. To accommodate complex scattering profiles, we also consider their linear evolution for normalized complex configuration weights. In this extension, normalization is retained, while a positive-probability interpretation is not imposed. Pomeron branching generates the linear growth term in the first-moment equation, while merging produces its quadratic nonlinearity. The full operator formulation can then be rewritten as a coherent-state path integral, providing a functional representation of the same microscopic dynamics \cite{Peliti:1985,Tauber:2005,Weber:2017}.

We derive the exact hierarchy of factorial correlators and obtain the effective evolution equations through a mean-field factorisation. In the selected network, the pomeron equation is autonomous, whereas the odderon evolves in the pomeron background through catalytic amplification and nonlinear fusion. This structure permits an analytical study of the competition between these mechanisms when diffusion is neglected. Its relation to the coupled evolution obtained in perturbative QCD is discussed at the level of the equations and their assumptions, without identifying the proposed reactions with a derivation from QCD \cite{Kovchegov:2004,Hatta:2005}.

The connection to elastic scattering is made by specifying the correspondence between the complex first correlators and the $C$-even and $C$-odd amplitude profiles. Once this correspondence, the initial profiles and the evolution parameters are fixed, the forward amplitude and the $\rho$ parameter follow from the evolved fields. A small odd component can modify the real part of the forward amplitude and produce differences between proton--proton and proton--antiproton scattering. The same decomposition determines the interference around a diffractive minimum. We illustrate these effects with simple initial profiles, without performing a fit to the scattering data.

%The goal of this paper is therefore twofold.

The objectives of this paper are twofold: First, we construct an
explicit reaction-based realisation of the complex logistic pomeron
equation together with a dynamically evolving odd sector. Second,
we examine how this evolution can affect elastic-scattering amplitudes
once a correspondence between the first moments and the amplitude
profiles is specified. The contribution lies in the selected
microscopic network, its exact operator and functional representations,
and the analytical study of its mean-field dynamics. The model
provides an effective microscopic origin for the evolution terms.

The paper is organized as follows: In Sec.~II, we introduce the reaction network, its master equations and the corresponding Doi--Peliti Hamiltonian. In Sec.~III, we derive the correlator hierarchy and its mean-field approximation. In Sec.~IV, we construct the coherent-state path integral and the effective action. In Sec.~V, we analyse the resulting evolution equations and the competition between catalytic amplification and fusion. In Sec.~VI, we establish the correspondence with scattering amplitudes and illustrate the consequences for $\rho$ and the dip region. In Sec.~VII, we present our conclusions and remarks. Explicit calculations of the master equations and operator relations are collected in the appendices A and B.

\section{Stochastic processes and configuration-weight evolution}
\label{sec:stochastic}

The Doi--Peliti formalism \cite{Doi:1976, Peliti:1985} provides an operator description of
reaction--diffusion processes, with applications ranging from chemical
kinetics to population dynamics \cite{Tauber:2005,Weber:2017}. Its starting point is a master equation
for occupation--number configurations: each reaction transfers weight
between configurations through a gain term and a corresponding loss
term. Representing these transfers with creation and annihilation
operators places the evolution in a bosonic Fock space and makes its
moment hierarchy accessible to field--theoretic methods.

We use this construction to investigate a reaction--based origin of the
complex nonlinear evolution introduced in our previous work on
high-energy hadronic scattering \cite{Kohara:2026}. The logistic form of that evolution
suggests a microscopic competition between branching and recombination.
Adding odderon excitations then allows the same construction to describe
an interacting pomeron--odderon system. Our aim is to derive the
Hamiltonian and the associated correlator hierarchy from local reaction
rules, and to identify the mean-field limit in which the pomeron sector
recovers the earlier equation.

The configuration weights in this representation are allowed to be
complex. The gain--loss evolution acts on these weights by the same linear
operator mapping, retaining its algebraic normalisation and Fock-space
structure.
For non-negative weights and reaction coefficients the construction has
the usual stochastic interpretation; complex weights provide the
ensemble used here to explore complex effective fields. The reaction
rules define the microscopic dynamics of this effective description,
with the pomeron and odderon treated as two species of excitations.
Rapidity $\tau$ parametrizes the evolution, and the two-dimensional
impact parameter $b$ specifies their transverse position.

We first discretize the transverse plane into cells indexed by $i$.
A configuration $\mathbf n$ specifies the number of pomerons  $n_{P,i}$ and odderons $n_{O,i}$ in each
cell on the lattice. Following the occupation-number representation
\cite{Doi:1976,Tauber:2005}, we associate with it a weight $\mathcal W(\mathbf n;\tau)$ and
write the general state together with its normalization:
\begin{equation}
 |\Phi(\tau)\rangle=\sum_{\mathbf n}\mathcal W(\mathbf n;\tau)
 |\mathbf n\rangle,\qquad
 \sum_{\mathbf n}\mathcal W(\mathbf n;\tau)=1.
 \label{eq:sec2:state}
\end{equation}
The expansion coefficients $\mathcal W$ carry the statistical weight
of each configuration. Their sum is fixed by the initial state and is
preserved by the gain--loss evolution. Allowing these coefficients to
be complex leaves the linear operator mapping unchanged. We stress that the same
algebraic cancellation also holds for complex reaction coefficients.
When such coefficients are used they are understood as effective
couplings rather than stochastic transition rates. These statements
assume that the configuration sums converge and that the evolution
is well defined on the states considered.

At each cell and for each $A=P,O$, the unnormalised occupation basis
obeys the creation and annihilation rules \cite{Doi:1976,Tauber:2005}

\begin{equation}
 \hat q_{A,i}|n_{A,i}\rangle=|n_{A,i}+1\rangle,\qquad
 \hat p_{A,i}|n_{A,i}\rangle=n_{A,i}|n_{A,i}-1\rangle,
\label{eq:sec2:operators}
\end{equation}
and the commutation relation
\begin{equation}
    [\hat p_{A,i},\hat q_{B,j}]=\delta_{AB}\delta_{ij}~.
\end{equation}
Writing the evolution generator as $-\hat H$ gives the
Schr\"odinger-like equation and its formal solution:
\begin{equation} \partial_\tau|\Phi\rangle=-\hat H|\Phi\rangle,\qquad
 |\Phi(\tau)\rangle=e^{-\hat H(\tau-\tau_0)}|\Phi(\tau_0)\rangle
 \label{eq:sec2:evolution}
\end{equation}
for a rapidity-independent Hamiltonian. 
The master equations determine the rapidity derivative of the state and their general expression for each reaction is given by
\begin{equation}
\partial_{\tau}\mathcal{W}(\mathbf n,\tau)=\sum_{\mathbf m\neq \mathbf n} \Big[w(\mathbf{n}|\mathbf{m})\,\mathcal{W}(\mathbf{m},\tau)-w(\mathbf{m}|\mathbf{n})\,\mathcal{W}(\mathbf{n},\tau)\Big]~, 
\label{eq:master:equation}
\end{equation}
where $w(\mathbf{m}|\mathbf{n})$ is the transition rate from a system in a initial configuration $\mathbf{n}$ to $\mathbf{m}$. The master equations together with the Schrodinger equation defines the Hamiltonian in terms of the considered reactions of a given model.
The relative sign between
the gain and loss terms fixes the sign of each contribution to
$\hat H$. In Appendix~\ref{AppendixA} we develop this correspondence
explicitly for every reaction in the model.

We select a local network combining pomeron branching and merging with
pomeron-catalysed odderon branching and ternary odderon fusion:
\begin{equation}
 P\xrightarrow{a_\ell}2P,\qquad
 2P\xrightarrow{r_\ell}P,\qquad
 P+O\xrightarrow{k_\ell}P+3O,\qquad
 3O\xrightarrow{u_\ell}O.
 \label{eq:sec2:channels}
\end{equation}

\begin{figure}[h]
    \centering
    \includegraphics[width=0.45\linewidth]{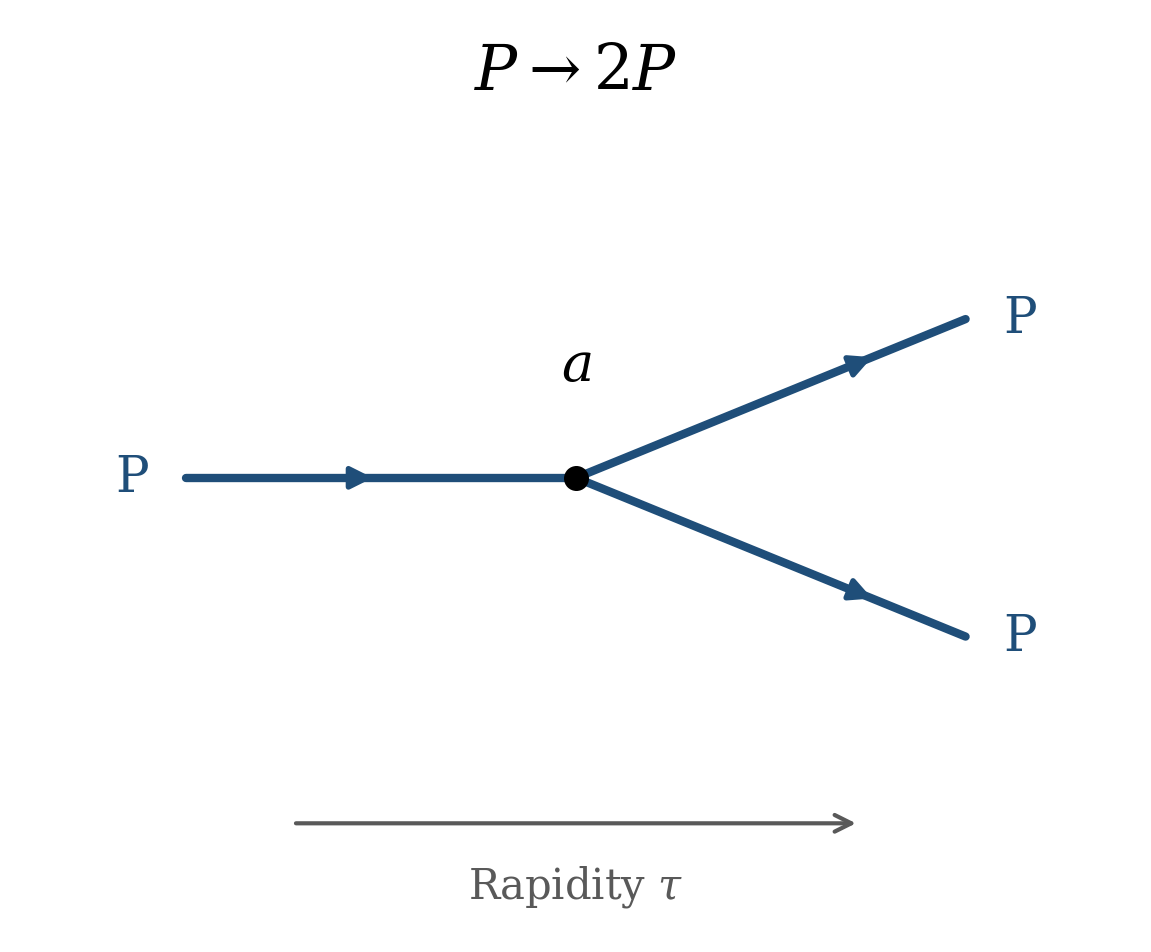}
    \includegraphics[width=0.45\linewidth]{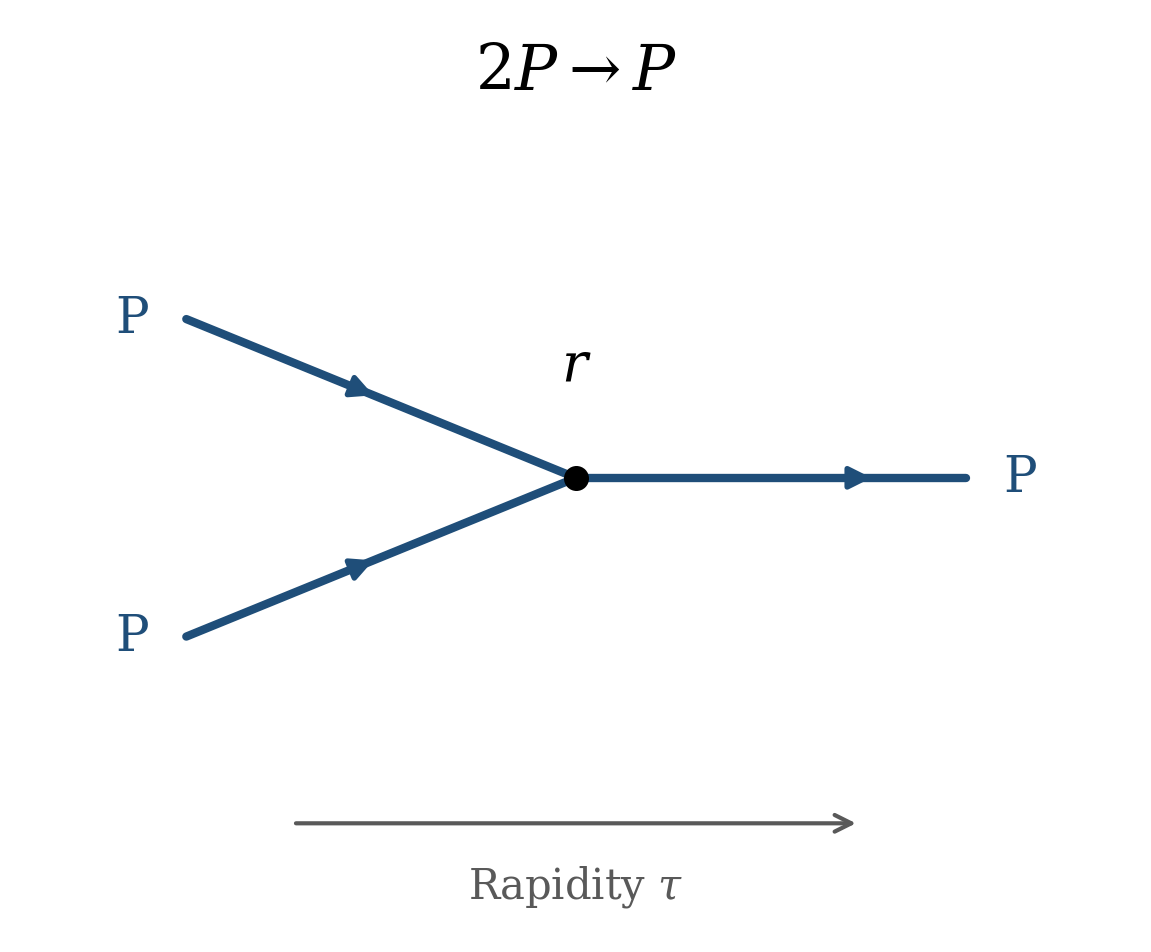}
    \includegraphics[width=0.45\linewidth]{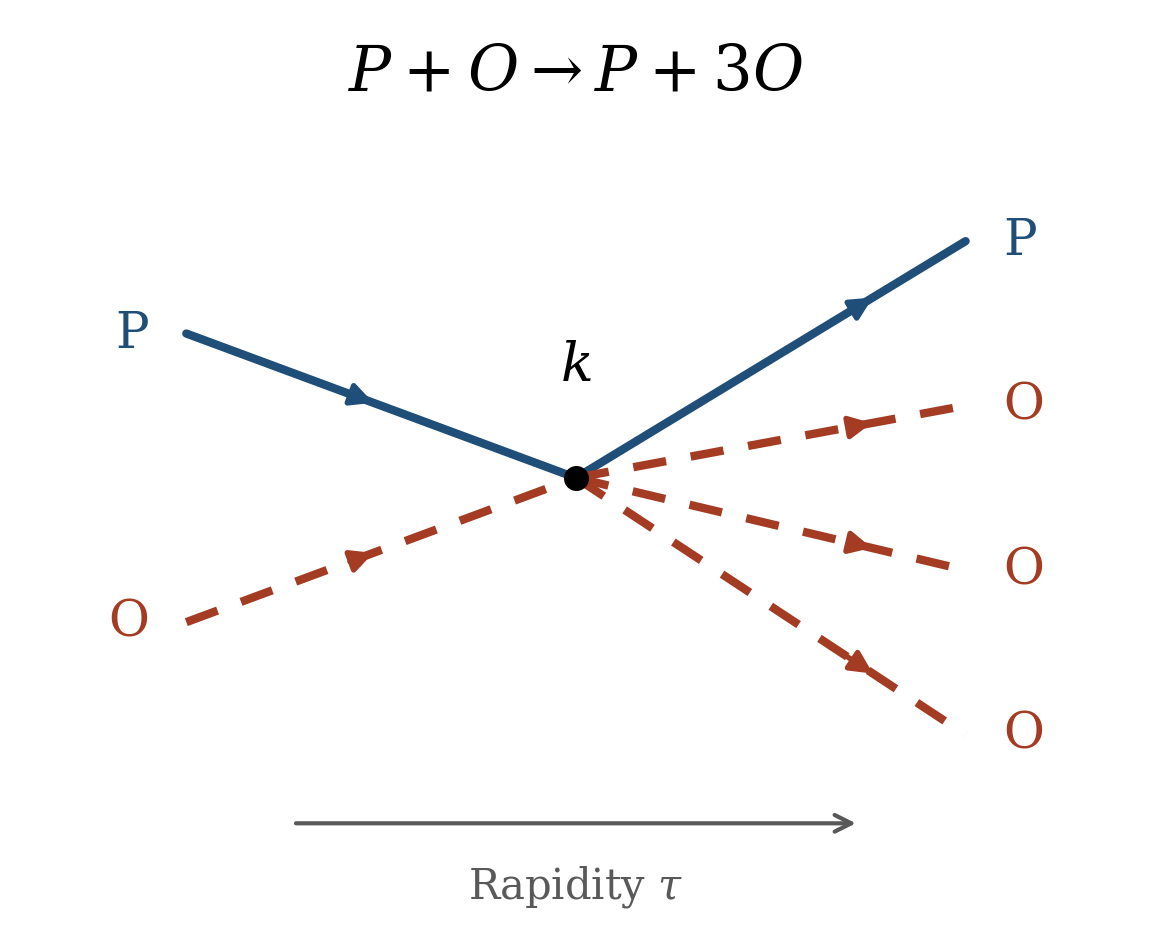}
    \includegraphics[width=0.45\linewidth]{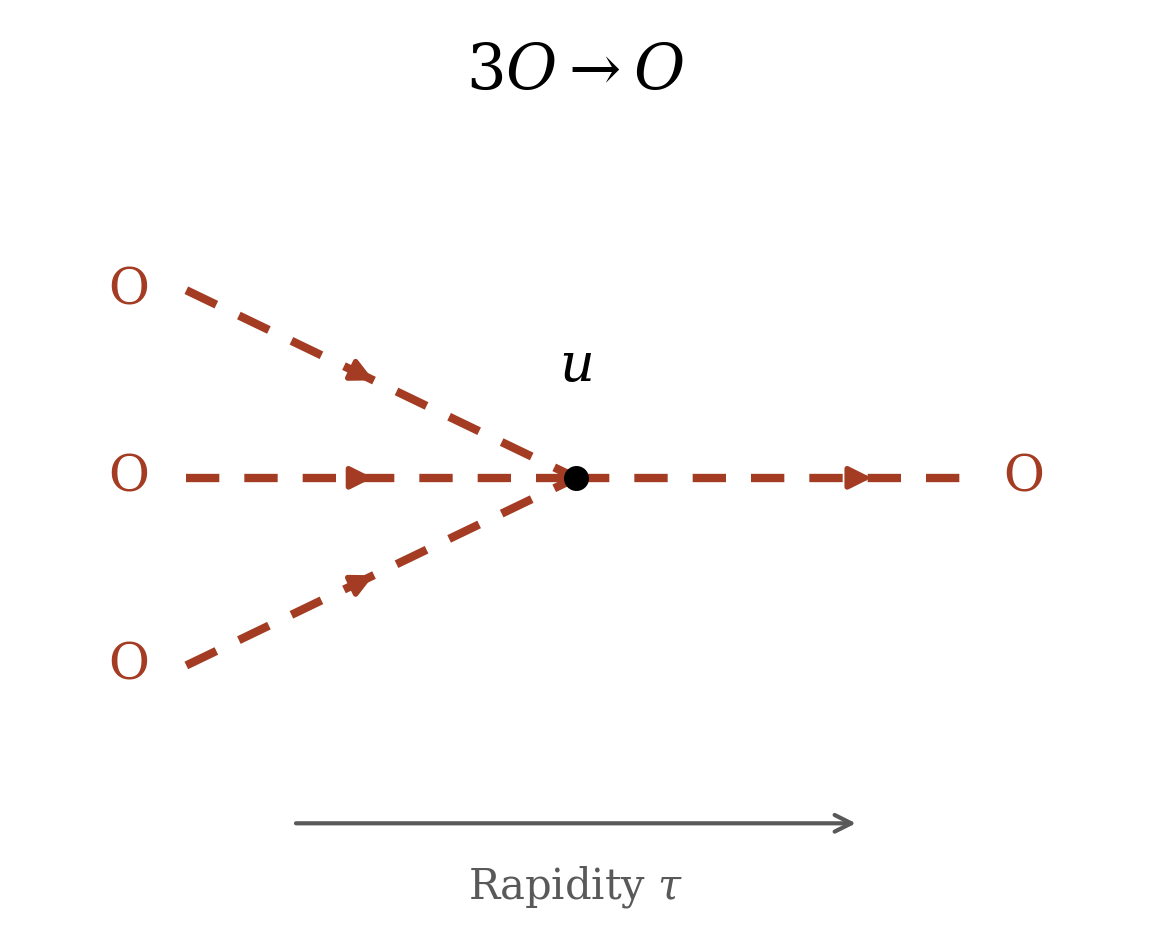}
    \caption{The figures show the minimum set of  channels of the microscopic reactions: pomeron branching and merging, pomeron-catalysed odderon branching and ternary odderon fusion.}
\label{fig:placeholder}
\end{figure}

Both species also hop between neighboring transverse cells. Each local
reaction conserves parity of the total odderon number. The network
is chosen so that its mean-field closure is covariant under
$P\mapsto P$, $O\mapsto-O$, as appropriate for the proposed effective
amplitude assignment. This is a selection criterion for the effective
model, not an assertion that these are all symmetry-allowed reactions.
The catalytic channel requires an odderon seed (an initial profile) and leaves the pomeron
number unchanged. Ternary fusion supplies nonlinear limitation of the
odderon field in the real positive mean-field regime.

The subscript $\ell$ denotes lattice reaction coefficients. Our convention
absorbs the identical-particle combinatorial factors into these coefficients.
The propensity of a reaction is its total rate of occurrence conditional
on the current configuration. At a cell with occupations $(n_P,n_O)$,
the four propensities are
$a_\ell n_P$, $r_\ell n_P(n_P-1)$, $k_\ell n_Pn_O$ and
$u_\ell n_O(n_O-1)(n_O-2)$. The falling factorials count the
ordered selections of reactants; the corresponding symmetry factors
are included in the coefficients. This convention is consistent with
the normally ordered operator construction \cite{Tauber:2005,Weber:2017}.
For complex coefficients, the same expressions are used algebraically
without a stochastic rate interpretation. Suppressing the cell index,
the reaction Hamiltonian is

\begin{eqnarray}
 \hat H_{\rm reac}&=&
 a_\ell(\hat q_P-\hat q_P^2)\hat p_P
 +r_\ell(\hat q_P^2-\hat q_P)\hat p_P^2\nonumber\\
 &&+k_\ell\hat q_P(\hat q_O-\hat q_O^3)\hat p_P\hat p_O
 +u_\ell(\hat q_O^3-\hat q_O)\hat p_O^3~.
 \label{eq:sec2:Hreac}
\end{eqnarray}
The full lattice Hamiltonian is the sum of these terms over cells plus
the hopping Hamiltonian, which is minus the generator derived in
Eq.~\eqref{EQ:Master_Equation_Diffusion}. With the sign convention in
Eq.~\eqref{eq:sec2:evolution}, its formal continuum diffusion term is
\begin{equation}
 \hat H_{\rm diff}=-\sum_{A=P,O}D_A\int d^2b\,
 \hat q_A(b)\nabla_b^2\hat p_A(b)~.
 \label{eq:sec2:Hdiff}
\end{equation}
The lattice supplies a spatial regulator for the local interactions.
The relation between the lattice coefficients and the continuous
fields is developed in Section~\ref{sec:hierarchy}, together with
the definition of coincident-point correlators.

The projection state from Doi-Peliti framework is defined as
\begin{equation}
\langle s|=\langle0|
\exp(\sum_{A,i}\hat p_{A,i})~,
\label{EQ:Projection:State:DP}
\end{equation}
with the property
\begin{equation}
   \langle s|\hat q_{A,i}=\langle s|~. \label{EQ:Projection:Q}
\end{equation}
 Thanks to the projection state, every term in
Eq.~\eqref{eq:sec2:Hreac}, as well as each hopping term, therefore
satisfies $\langle s|\hat H=0$. Consequently,
\begin{equation}
 \partial_\tau\sum_{\mathbf n}\mathcal W(\mathbf n;\tau)
 =-\langle s|\hat H|\Phi(\tau)\rangle=0.
 \label{eq:sec2:conservation}
\end{equation}
Conservation of the total weight is therefore built into the
Hamiltonian. Particle numbers nevertheless change under branching,
recombination, and their moments carry the dynamical
information of the model. We now derive the equations governing those
moments and their relation to continuous correlators.

\section{Hierarchy of correlators}
\label{sec:hierarchy}

\subsection{Projection state and weighted expectations}
To extract moments from the Fock-space state, we use the projection
bra that sums its configuration coefficients. On the transverse
lattice this bra takes the form
\begin{equation}
 \langle s|=\langle0|\exp\left(\sum_{A=P,O}\sum_i\hat p_{A,i}\right).
 \label{EQ:BS:def}
\end{equation}
The unit coefficient of each annihilation operator in the exponential
ensures that every occupation-number state has unit overlap with
$\langle s|$. Using the commutation relations then gives
\begin{equation}
 \langle s|\hat q_{A,i}=\langle s|,\qquad
 \langle s|\mathbf n\rangle=1,\qquad
 \langle s|\Phi(\tau)\rangle=\sum_{\mathbf n}\mathcal W(\mathbf n;\tau)=1.
 \label{EQ:q_applied_to_sum}
\end{equation}
Thus $\langle s|\Phi\rangle$ gives the total configuration weight.
The coefficient of a specified configuration follows from its overlap
with the corresponding occupation-number bra. Since
$\langle\mathbf n|\mathbf m\rangle=
\delta_{\mathbf n,\mathbf m}\prod_{A,i}n_{A,i}!$
in the unnormalised basis, this coefficient is
\begin{equation}
 \mathcal W(\mathbf n;\tau)
 =\frac{\langle\mathbf n|\Phi(\tau)\rangle}
 {\prod_{A,i}n_{A,i}!}.
 \label{EQ:configuration_weight_extraction}
\end{equation}
For an arbitrary function $F(\mathbf n)$ of the occupation numbers, the weighted
expectation is
\begin{equation}
 \langle F\rangle_\tau
 =\sum_{\mathbf n}F(\mathbf n)\mathcal W(\mathbf n;\tau)
 =\langle s|F(\{\hat q_{A,i}\hat p_{A,i}\})|\Phi(\tau)\rangle.
 \label{EQ:observable_coherent}
\end{equation}
This weighted expectation provides the same operator identities for
real and complex ensembles. Products of annihilation operators select
falling factorials of the occupation numbers; in particular,
$\langle s|\hat p_{A,i}^n|\Phi\rangle=\langle(n_{A,i})_n\rangle$
is a factorial moment, with $(n)_k=n(n-1)\cdots(n-k+1)$.

For a possibly rapidity-dependent operator $\hat F$, differentiating
its expectation gives
\begin{align}
 \partial_\tau\langle\hat F\rangle
 &=\langle s|\partial_\tau\hat F|\Phi\rangle
   -\langle s|\hat F\hat H|\Phi\rangle\nonumber\\
 &=\langle\partial_\tau\hat F\rangle
   +\langle s|[\hat H,\hat F]|\Phi\rangle,
 \label{EQ:Evolution_Equation_Observable}
\end{align}
because the conservative generator satisfies $\langle s|\hat H=0$.
For an operator without explicit rapidity dependence the average of the Heisenberg-like equation is
\begin{equation}
 \partial_\tau\langle\hat F\rangle
 =\langle[\hat H,\hat F]\rangle.
 \label{EQ:Evolution_Equation_Observable_1}
\end{equation}
Equation~\eqref{EQ:Evolution_Equation_Observable_1} expresses the
master-equation dynamics directly in terms of observable operators.
It is the starting point for the correlator hierarchy.

\subsection{Lattice moments and continuum correlators}
The first moment on a lattice cell is the weighted occupation moment:
\begin{equation}
 g_{A,i}(\tau)=\langle s|\hat p_{A,i}|\Phi(\tau)\rangle =\langle n_{A,i}\rangle
 \label{eq:hier:lattice_first}
\end{equation}
To introduce a field per unit transverse area, let
$v=a_\perp^2$ be the area of a square cell. Dividing the annihilation
operator by this area and retaining the creation operator gives
\begin{equation}
 \hat p_A(b_i)=\hat p_{A,i}/v,\qquad
 \hat q_A(b_i)=\hat q_{A,i}.
 \label{eq:hier:field_scaling}
\end{equation}
Then $\delta_{ij}/v$ tends to $\delta^{(2)}(b-b')$, giving the commutation relation
\begin{equation}
 [\hat p_A(b),\hat q_B(b')]=\delta_{AB}\delta^{(2)}(b-b')~.
 \label{EQ:comutator}
\end{equation}
The projection bra state has the formal continuous representation
\begin{equation}
 \langle s|=\langle0|\exp\left(\sum_{A=P,O} \int d^2b\,\hat p_A(b)\right).
 \label{eq:hier:continuous_projection}
\end{equation}
In this convention the continuum one-point function is
\begin{equation}
 G_A(b,\tau)=\langle s|\hat p_A(b)|\Phi(\tau)\rangle
 =\lim_{a_\perp\to0}\frac{g_{A,i}(\tau)}{v},\qquad b_i\to b,
 \label{eq:hier:G_one}
\end{equation}
with the limit understood through the lattice regularisation. This
normalisation gives $G_A$ the dimensions of an areal density:
$\sum_i g_{A,i}=\sum_i vG_A(b_i)$ becomes
$\int d^2b\,G_A(b)$ in the continuous description. A positive-density
interpretation requires positive probability weights and complex $G_A$ are
weighted fields with the same spatial normalisation.

From the appendix A it is easy to see that with $m$ incoming excitations, the propensity of a local reaction term has $m$
annihilation operators. Under Eq.~\eqref{eq:hier:field_scaling},
its continuous coefficient is
$\lambda^{\rm cont}=v^{m-1}\lambda^{\rm lat}$.
For the retained channels this gives
\begin{equation}
 a=a_\ell,\qquad r=vr_\ell,\qquad k=vk_\ell,\qquad u=v^2u_\ell.
 \label{eq:hier:rate_scaling}
\end{equation}
We use these continuous coefficients in the correlator equations and
omit the superscripts once the convention is established. Local
products retain their lattice regularisation, which also specifies
the coincident-point limits appearing in the reaction terms.

For arbitrary species labels, the equal-rapidity factorial correlator is
\begin{equation}
 G_{A_1\cdots A_n}(b_1,\ldots,b_n;\tau)
 =\left\langle\hat p_{A_1}(b_1)\cdots\hat p_{A_n}(b_n)\right\rangle_\tau.
 \label{eq:hier:general_correlator}
\end{equation}
Examples are $G_{PP}(b_1,b_2)$, $G_{OO}(b_1,b_2)$,
$G_{PO}(b_1,b_2)$ and $G_{OOO}(b_1,b_2,b_3)$.
At the same lattice site these are scaled factorial moments:
$G_{PP}(b_i,b_i)=\langle n_{P,i}(n_{P,i}-1)\rangle/v^2$.
Their departure from products of first moments measures the
factorial correlations generated by the interacting evolution.

Creation operators placed to the left in a normal-ordered expectation
can be removed by the projection property. For example,
$\langle\hat q_A(b)\hat p_B(b')\rangle=G_B(b')$.
Reversing their order produces a contact term:
\begin{equation}
 \langle\hat p_A(b)\hat q_B(b')\rangle
 =G_A(b)+\delta_{AB}\delta^{(2)}(b-b').
 \label{eq:hier:pq_contact}
\end{equation}
This ordering relation accounts for the contact contribution.
The mixed factorial correlator $G_{PO}$, in contrast, describes the
joint occupation of the two species through two annihilation
operators. At coincident positions the same lattice prescription
fixes both quantities.

\subsection{Exact first-moment equations}
For a normal-ordered local reaction Hamiltonian, the commutation
relations imply
\begin{equation}
 [\hat H_{\rm reac},\hat p_{A,i}]
 =-\frac{\partial\hat H_{\rm reac}}{\partial\hat q_{A,i}},
 \label{eq:hier:derivative_rule}
\end{equation}
where the derivative is an algebraic derivative of the normal-ordered
polynomial. Projection then replaces the creation operators by one.
At one cell, projection gives
\begin{align}
 \langle[H_{\rm reac},\hat p_P]\rangle
&=a_\ell\langle\hat p_P\rangle-r_\ell\langle\hat p_P^2\rangle, \label{eq:hier:projected_commutators:Even}\\
 \langle[H_{\rm reac},\hat p_O]\rangle
 &=2k_\ell\langle\hat p_P\hat p_O\rangle-2u_\ell\langle\hat p_O^3\rangle.
\label{eq:hier:projected_commutators}
\end{align}
The channel-by-channel commutators are evaluated in
Appendix~\ref{Appendix_B}.
In the continuum diffusion commutator, the delta function is integrated
against the local Hamiltonian density:
\begin{equation}
 [H_{{\rm diff},A},\hat p_A(b)]
 =D_A\int d^2b'\,\delta^{(2)}(b'-b)\nabla_{b'}^2\hat p_A(b')
 =D_A\nabla_b^2\hat p_A(b).
 \label{eq:hier:diffusion_commutator}
\end{equation}
The integration selects the position of the observed field. Local
reaction terms follow the same procedure and yield the correlators
evaluated at that position.

Consequently the exact, unclosed continuum equations are
\begin{align}
 \partial_\tau G_P(b)&=D_P\nabla_b^2G_P(b)+aG_P(b)-rG_{PP}(b,b),
 \label{EQ:Evolution_Equation_Correlator_Pomeron}\\
 \partial_\tau G_O(b)&=D_O\nabla_b^2G_O(b)+2kG_{PO}(b,b)-2uG_{OOO}(b,b,b).
 \label{EQ:Evolution_Equation_Correlator_Odderon}
\end{align}
The common rapidity argument is suppressed. The first moments depend
on a second factorial moment and a mixed second moment, while ternary
fusion introduces a third factorial moment. Applying the same evolution
identity to these observables generates the higher hierarchy similar to RFT \cite{Alessandrini:1977} or JIMWLK \cite{
Balitsky:1996,Weigert:2005,Iancu:2001} hierarchies. For example,
at one cell the pomeron moments $m_j=\langle\hat p_P^j\rangle$ satisfy
\begin{align}
 \dot m_1&=a_\ell m_1-r_\ell m_2, \label{eq:hier:example_first_moment}\\
 \dot m_2&=2a_\ell(m_2+m_1)-2r_\ell(m_3+m_2).
 \label{eq:hier:example_second_moment}
\end{align}
The general hierarchy and its explicit mixed-species examples are
provided in Appendix~\ref{Appendix_B}.

\subsection{Mean-field closure and its covariance}
The mean-field approximation neglects the local factorial cumulants
entering the reaction terms. The multi-point correlation functions factorizes as
\begin{equation}
 G_{PP}(b,b)\simeq G_P(b)^2,\qquad
 G_{PO}(b,b)\simeq G_P(b)G_O(b),\qquad
 G_{OOO}(b,b,b)\simeq G_O(b)^3.
 \label{eq:hier:closure}
\end{equation}
These are the corresponding moments of independent local Poisson
weights. Adopting them as a closure does not imply that the exact
interacting state remains Poissonian. The same algebraic factorization
is used for the complex weighted ensemble. With $P=G_P$ and $O=G_O$ for the pomeron and odderon respectively,
the first-moment equations become
\begin{eqnarray}
 \partial_\tau P&=&D_P\nabla_b^2P+aP-rP^2,
 \label{eq:sec2:meanP}\\
 \partial_\tau O&=&D_O\nabla_b^2O+2kPO-2uO^3.
 \label{eq:sec2:meanO}
\end{eqnarray}
pomeron branching and merging produce the linear growth and quadratic
limitation terms. The catalytic process contributes $2kPO$ without
changing pomeron occupations, and ternary fusion contributes $-2uO^3$.
The pomeron marginal is autonomous already in the microscopic dynamics;
its logistic equation additionally requires the factorial closure.
It is therefore not necessary to neglect a feedback term to obtain
the pomeron background in this model.

With the four chosen reactions the closed equations are covariant under
\begin{equation}
 P\mapsto P,\qquad O\mapsto-O.
 \label{eq:hier:field_covariance}
\end{equation}
At fixed $P$, opposite odderon seeds evolve into opposite fields.
This supports the effective assignment of even and odd exchange
profiles used in the later amplitude matching. It is a property of
the mean-field closure, distinct from conservation of number parity
in the exact Fock-space dynamics. Appendix~\ref{Appendix_B} gives
the explicit distinction at the generating-function level.

For $D_P=0$, the pomeron equation is
\begin{equation}
 \partial_\tau G_P=aG_P-rG_P^2.
 \label{eq:hier:decoupled_logistic}
\end{equation}
In the convention $T_+=i\mathcal N_PG_P$, with a real positive
normalization $\mathcal N_P$, it yields
\begin{equation}
\partial_\tau T_+=aT_++i(r/\mathcal N_P)T_+^2~,
\label{EQ:Logistic:Pomeron}
\end{equation}
recovering the
preceding effective evolution equation \cite{Kohara:2026}.
The complex profiles, their real positive saturation regime, and
the interpretation of the odd component are analysed after the
functional formulation.

\section{From the Hamiltonian to a functional integral}
\label{sec:path_integral}

The operator evolution can be expressed as a functional integral by
resolving the propagator into coherent states at successive rapidities \cite{Peliti:1985,Wiese:2016}.
This construction retains the reaction Hamiltonian and the initial
configuration weights, and provides an action whose field correlators
represent the factorial correlators introduced in
Section~\ref{sec:hierarchy}. We first keep the transverse lattice and
the rapidity discretisation, which together specify the operator
ordering and the endpoint factors of the functional integral.

\subsection{Coherent states and rapidity discretisation}
We define for both species at all lattice cells the unnormalised
coherent ket and its dual by \cite{Peliti:1985,Wiese:2016}

\begin{equation}
 |\boldsymbol\phi\rangle
 =\exp\left(\sum_{A,i}\phi_{A,i}\hat q_{A,i}\right)|0\rangle,
 \qquad
 \langle\bar{\boldsymbol\phi}|
 =\langle0|\exp\left(\sum_{A,i}\bar\phi_{A,i}\hat p_{A,i}\right).
 \label{eq:path:coherent_states}
\end{equation}
They satisfy the eigenvalue relations
\begin{align}
 \hat p_{A,i}|\boldsymbol\phi\rangle
 &=\phi_{A,i}|\boldsymbol\phi\rangle,&
 \langle\bar{\boldsymbol\phi}|\hat q_{A,i}  =\bar\phi_{A,i}\langle\bar{\boldsymbol\phi}|
 ~,
 \label{eq:path:eigenvalues}
\end{align}
with the scalar product
\begin{equation}
 \langle\bar{\boldsymbol\phi}|\boldsymbol\psi\rangle
 =\exp\left(\sum_{A,i}\bar\phi_{A,i}\psi_{A,i}\right).
\end{equation}
The coherent-state resolution of the identity is
\begin{equation}
 \mathbf 1=\int\prod_{A,i}\frac{d^2\phi_{A,i}}{\pi}
 \exp\left(-\sum_{A,i}|\phi_{A,i}|^2\right)
 |\boldsymbol\phi\rangle\langle\boldsymbol\phi^*|.
 \label{eq:path:identity}
\end{equation}
In Eq.~\eqref{eq:path:identity}, the dual parameter lies on the original
contour $\bar\phi=\phi^*$. The analytic normal symbol is nevertheless
written as a function of two field arguments. Stationary equations are
obtained by varying these arguments independently in the complexified
representation; their saddles need not satisfy $\bar\phi=\phi^*$.
Any contour deformation used to reach such a saddle must respect
convergence and avoid singularities. The bar labels the dual coherent
parameter rather than a conjugacy condition imposed on a saddle.

We divide $[\tau_0,\tau_f]$ into $N$ intervals of length $\Delta\tau$.
For a normal-ordered Hamiltonian, its short-step kernel is
\begin{align}
 \langle\bar{\boldsymbol\phi}_{\ell}|
 e^{-\Delta\tau\hat H}|\boldsymbol\phi_{\ell-1}\rangle
 ={}&\exp\left(\bar{\boldsymbol\phi}_{\ell}\cdot\boldsymbol\phi_{\ell-1}\right)
 \big[1-\Delta\tau\,\mathcal H(\bar{\boldsymbol\phi}_{\ell},\boldsymbol\phi_{\ell-1})    +O(\Delta\tau^2)\big].
\label{eq:path:short_step}
\end{align}
Here $\mathcal H$ is the normal symbol of $\hat H$, obtained by
$\hat q\mapsto\bar\phi$ and $\hat p\mapsto\phi$. The density
variables in each reaction are evaluated at the preceding rapidity
slice, while the creation variables are evaluated at the following
slice. This ordering supplies the forward prescription. Retaining the
discrete ordering is important when products of fields at coincident
rapidities are considered \cite{Andreanov:2006}.

The initial state enters through its generating function
\begin{equation}
 \mathcal Z_0(\bar{\boldsymbol\phi})
 =\langle\bar{\boldsymbol\phi}|\Phi_0\rangle
 =\sum_{\mathbf n}\mathcal W_0(\mathbf n)
   \prod_{A,i}\bar\phi_{A,i}^{n_{A,i}}.
 \label{eq:path:initial_generating_function}
\end{equation}
At the final endpoint the projection state gives
$\langle s|\boldsymbol\phi_N\rangle=
\exp(\sum_{A,i}\phi_{A,i,N})$. Inserting the coherent-state identities
therefore yields the normalisation functional
\begin{eqnarray}
 \mathcal Z &=&\langle s|e^{-(\tau_f-\tau_0)\hat H}|\Phi_0\rangle
 \nonumber\\
 &=&\lim_{N\to\infty}\int\prod_{\ell=0}^{N}d\nu(\boldsymbol\phi_\ell)
 \;\mathcal Z_0(\bar{\boldsymbol\phi}_0)
 \exp\Bigg[-\sum_{\ell=1}^{N}
\bar{\boldsymbol\phi}_{\ell}\cdot
(\boldsymbol\phi_\ell-\boldsymbol\phi_{\ell-1})
 \nonumber\\
 &&\hspace{18mm}-\bar{\boldsymbol\phi}_0\cdot\boldsymbol\phi_0
+\mathbf1\cdot\boldsymbol\phi_N
       -\Delta\tau\sum_{\ell=1}^{N}
       \mathcal H(\bar{\boldsymbol\phi}_\ell, \boldsymbol\phi_{\ell-1})\Bigg],
 \label{eq:path:discrete_integral}
\end{eqnarray}
where $d\nu(\boldsymbol\phi)=\prod_{A,i}d^2\phi_{A,i}/\pi$ denotes
the contour measure, with its Gaussian factor already displayed in
the exponent. For a normalised initial state, $\mathcal Z=1$.
Sources or observable insertions produce the nontrivial generating
functionals and correlators; $\mathcal Z$ itself is the conserved
normalisation rather than a thermal partition function  \cite{Weber:2017}.

\subsection{Action and displacement of the response field}
The overlap between adjacent coherent states produces the kinetic
term $\bar\phi\,\partial_\tau\phi$, and the evolution operator
$e^{-\Delta\tau H}$ supplies $+\mathcal H$ in the action. Keeping the
continuum spatial normalisation of Section~\ref{sec:hierarchy}, the
bulk expression before displacement is
\begin{equation}
 S_{\rm bulk}[\bar\phi,\phi]=
 \int_{\tau_0}^{\tau_f}d\tau\int d^2b\,
\left[\sum_{A=P,O}\bar\phi_A(\partial_\tau-D_A\nabla_b^2)\phi_A+\mathcal H_{\rm reac}(\bar\phi,\phi)\right].
 \label{eq:path:unshifted_action}
\end{equation}
Its endpoint prescription is fixed by
Eq.~\eqref{eq:path:discrete_integral}. The field $\phi_A$ represents
annihilation-operator insertions, and $\bar\phi_A$ represents the
creation-operator symbol. This assignment preserves the original
operator notation: $\hat p_A\mapsto\phi_A$ and
$\hat q_A\mapsto\bar\phi_A$.

The projection property singles out $\bar\phi_A=1$. We therefore
introduce the displaced Doi-response fields \cite{Peliti:1985,Weber:2017,Wiese:2016}
\begin{equation}
 \bar\phi_A=1+\widetilde\phi_A.
 \label{eq:path:doi_shift}
\end{equation}
In Eq.~\eqref{eq:path:discrete_integral}, the constant part of the
kinetic sum telescopes against the final projection factor, as in the
standard treatment of the Doi shift \cite{Wiese:2016}. The
remaining endpoint factor and discrete bulk action are
\begin{align}
 \mathcal Z={}&\lim_{N\to\infty}\int\mathcal D_N\phi\,      \mathcal D_N\widetilde\phi\;
 \mathcal Z_0(\mathbf1+\widetilde{\boldsymbol\phi}_0)
 \exp\left[-\widetilde{\boldsymbol\phi}_0\cdot\boldsymbol\phi_0 -S_N\right],\nonumber\\
 S_N={}&\sum_{\ell=1}^{N}\widetilde{\boldsymbol\phi}_\ell\cdot(\boldsymbol\phi_\ell-\boldsymbol\phi_{\ell-1})
 +\Delta\tau\sum_{\ell=1}^{N}
 \mathcal H(\mathbf1+\widetilde{\boldsymbol\phi}_\ell,\boldsymbol\phi_{\ell-1}).
\label{eq:path:shifted_discrete_integral}
\end{align}
The displacement also shifts the integration contours. Equations
\eqref{eq:path:discrete_integral} and
\eqref{eq:path:shifted_discrete_integral} specify the same functional
integral; the bulk continuum action is compact notation for this
prescription. Under spatial boundary conditions that remove the
integral of $\nabla_b^2\phi_A$, it is
\begin{equation}
 S_{\rm bulk}[\widetilde\phi,\phi]=
 \int_{\tau_0}^{\tau_f}d\tau\int d^2b\, \left[\sum_A(\widetilde\phi_A
\partial_\tau-D_A\nabla_b^2)\phi_A+\mathcal V(\widetilde\phi,\phi)\right].
 \label{eq:path:shifted_action}
\end{equation}
With the continuum coefficients of Section~\ref{sec:hierarchy},
the reaction polynomial is
\begin{eqnarray}
 \mathcal V &=&-a(\widetilde\phi_P+\widetilde\phi_P^2)\phi_P+r(\widetilde\phi_P+\widetilde\phi_P^2)\phi_P^2\nonumber\\
 &&-k(1+\widetilde\phi_P)
(2\widetilde\phi_O+3\widetilde\phi_O^2+\widetilde\phi_O^3) \phi_P\phi_O\nonumber\\
&&+u(2\widetilde\phi_O+3\widetilde\phi_O^2+\widetilde\phi_O^3)
\phi_O^3.
\label{eq:path:full_polynomial}
\end{eqnarray}
Each monomial contains a response field, and
$\mathcal V(0,\phi)=0$ expresses conservation of the total weight.
The factor $1+\widetilde\phi_P$ in the catalytic term retains the
creation operator restoring the catalyst. The higher response powers
come from the full gain--loss polynomial, not from a truncation to
the mean-field drift. This distinction between the full reaction action
and its deterministic approximation is also important in single-species
reaction theories

The initial ensemble remains part of the functional representation.
For independent Poisson weights with cell means $g_{A,i}^{(0)}$, the initial functional in the displaced response field is
\begin{equation}
 \mathcal Z_0(\mathbf1+\widetilde{\boldsymbol\phi}_0)
 =\exp\left(\sum_{A,i}g_{A,i}^{(0)}\widetilde\phi_{A,i,0}\right).
 \label{eq:path:poisson_initial}
\end{equation}
At the density level this produces an initial factor with
$g_{A,i}^{(0)}=v\rho_A^{(0)}(b_i)$. The boundary saddle then sets
$\phi_A(\tau_0,b)=\rho_A^{(0)}(b)$, while the final projection
sets $\widetilde\phi_A(\tau_f,b)=0$. A specified occupation state
instead has $\mathcal Z_0(\bar\phi)=\prod_{A,i}\bar\phi_{A,i}^{n_{A,i}^{(0)}}$;
correlated initial weights are encoded by the corresponding joint
generating function. These initial factors also encode the initial
factorial cumulants. Complex configuration weights enter through the
same generating function, within the convergent or analytically
continued domain of the coherent-state representation.

\subsection{Functional correlators and the saddle correspondence}
Normal-ordered annihilation products have the functional insertions
\begin{equation}
 G_{A_1\cdots A_n}(b_1,\ldots,b_n;\tau)
 =\left\langle\phi_{A_1}(b_1,\tau)\cdots   \phi_{A_n}(b_n,\tau)\right\rangle_{S,\mathcal Z_0},
 \label{eq:path:factorial_insertions}
\end{equation}
where the expectation includes the endpoint factors in
Eq.~\eqref{eq:path:shifted_discrete_integral}. The annihilation field thus
represents the factorial observables of the operator theory; the
functional construction does not require the evolving Fock-space
state to remain a coherent state. In particular, $\phi_A$ is an
annihilation-field variable representing factorial observables. It need
not be a microscopic occupation density in each field configuration
\cite{Andreanov:2006}.

Varying Eq.~\eqref{eq:path:shifted_action} with respect to the response
field and evaluating on the zero-response saddle gives  the
mean-field equation \cite{Elgart:2004,Tauber:2005}

\begin{equation}
 (\partial_\tau-D_A\nabla_b^2)\phi_A
 =-\left.\frac{\partial\mathcal V}{\partial\widetilde\phi_A}\right|_{\widetilde\phi=0}.
 \label{eq:path:saddle_rule}
\end{equation}
The two response derivatives are
\begin{equation}
 \left.\partial_{\widetilde\phi_P}\mathcal V\right|_0
 =-a\phi_P+r\phi_P^2,
 \qquad
 \left.\partial_{\widetilde\phi_O}\mathcal V\right|_0
 =-2k\phi_P\phi_O+2u\phi_O^3.
 \label{eq:path:response_derivatives}
\end{equation}
The zero-response saddle therefore reproduces
Eqs.~\eqref{eq:sec2:meanP}--\eqref{eq:sec2:meanO}, with
$P\mapsto\phi_P$ and $O\mapsto\phi_O$.
This recovers the closure already introduced in
Section~\ref{sec:hierarchy}, rather than introducing a second
approximation to the exact operator hierarchy. Identifying the
first moments with these saddle fields neglects fluctuation
contributions to the higher correlators. The response vertices
provide the functional framework for corrections to that closure.

\subsection{Quadratic action and retarded propagation}
Around zero annihilation and zero response fields, the bulk quadratic
action is
\begin{equation}
 S_0=\int d\tau\,d^2b\;
 \widetilde{\boldsymbol\phi}^{\,T}
 [\mathbf1\partial_\tau-\mathbf D\nabla_b^2-\mathbf M]
 \boldsymbol\phi,
 \qquad
 \mathbf D=\begin{pmatrix}D_P&0\\0&D_O\end{pmatrix},\quad
 \mathbf M=\begin{pmatrix}a&0\\0&0\end{pmatrix}.
 \label{eq:path:quadratic_matrix}
\end{equation}
Both mixed catalytic drift and ternary fusion are interaction terms
in this expansion. They do not contribute off-diagonal bare propagation
or an independent linear odderon growth coefficient.

The bare response propagator is the connected mixed two-point function
\begin{equation}
 R_{AB}(\tau,b;\tau',b')
 =\langle\phi_A(\tau,b)\widetilde\phi_B(\tau',b')\rangle_{0,c}.
 \label{eq:path:response_definition}
\end{equation}
It measures the linear response of $\phi_A$ to a source $j_B$
introduced through $S_0\mapsto S_0-\int\widetilde\phi_B j_B$.
It is thus a two-point function of an annihilation field and a response
field, distinct from the factorial field correlator
$\langle\phi_A\phi_B\rangle$. Retardation means that it vanishes
for $\tau<\tau'$ \cite{Tauber:2005}.

For constant coefficients, its equation and transverse Fourier
representation are
\begin{align}
 [\mathbf1\partial_\tau-\mathbf D\nabla_b^2-\mathbf M]\mathbf R
 &=\mathbf1\delta(\tau-\tau')\delta^{(2)}(b-b'),\nonumber\\
 \mathbf R(\boldsymbol{l},\Delta\tau)
 &=\theta(\Delta\tau)
 \begin{pmatrix}
 e^{(a-D_P\,l^2)\Delta\tau}&0\\
 0&e^{-D_O\,l^2\Delta\tau}
 \end{pmatrix},
 \label{eq:path:matrix_response}
\end{align}
where $\Delta\tau=\tau-\tau'$ and $l^2=|\boldsymbol l|^2$ is the square of transferred momentum.
The forward-slice prescription sets the equal-slice contraction to
zero, conventionally written as $\theta(0)=0$. This value specifies
the causal contraction convention inherited from normal ordering,
rather than an independent choice for the continuum kernel
\cite{Andreanov:2006,Tauber:2005}. For $D_A>0$, we obtain
\begin{equation}
K_A(\Delta\tau,\Delta b)=
\frac{\theta(\Delta\tau)}{4\pi D_A\Delta\tau}\,
 \exp\!\left(-\frac{|\Delta b|^2}{4D_A\Delta\tau}           +m_A\Delta\tau\right)~ ,
\label{eq:path:matrix_heat_kernel}
\end{equation}
with $m_P=a,\quad m_O=0$.
Then $R_{AB}=\delta_{AB}K_A$. In particular,
\begin{equation}
 R_P=\theta(\Delta\tau)\frac{
 e^{-|\Delta b|^2/(4D_P\Delta\tau)+a\Delta\tau}}
 {4\pi D_P\Delta\tau}.
 \label{eq:path:P_heat_kernel}
\end{equation}
The limit $D_P=0$ is
\begin{equation}
R_P=\theta(\Delta\tau)\,e^{m_A\Delta\tau}\,\delta^{(2)}(\Delta b)~.    
\end{equation}
A nonzero pomeron background can amplify the odderon through the
catalytic interaction even though the bare odderon propagator has
no exponential growth factor. The kernels above refer specifically
to the expansion about zero fields. They are not the response
operators linearised around a finite evolving background.

\subsection{Reaction vertices}
Writing $S_{\rm bulk}=S_0+S_{\rm int}$ separates bare propagation from
vertices with three or more fields. In the single-species pomeron
limit the action takes the explicit form
\begin{equation}
 S_P=\int d\tau\,d^2b\;
 \left[\widetilde\phi_P(\partial_\tau-D_P\nabla_b^2-a)\phi_P
       -a\widetilde\phi_P^2\phi_P
       +r\widetilde\phi_P\phi_P^2
       +r\widetilde\phi_P^2\phi_P^2\right],
 \label{eq:path:P_action}
\end{equation}
with the initial factors retained as above. Branching therefore
contributes both the linear growth operator and a cubic vertex;
recombination contributes a cubic and a quartic vertex. The catalytic and ternary-fusion processes generate the remaining
monomials displayed in Eq.~\eqref{eq:path:full_polynomial}.

Perturbative contractions join a density field at a later rapidity to
a response field at an earlier rapidity through $R_{AB}$. For a term
$g\,\widetilde\phi_P^{r_P}\widetilde\phi_O^{r_O}
\phi_P^{s_P}\phi_O^{s_O}$ in the action, expansion of $e^{-S_{\rm int}}$
produces the coefficient $-g$ together with its Wick multiplicities.
If an amputated-vertex convention includes factorials for identical
legs, its vertex factor is
$-g\,r_P!r_O!s_P!s_O!$. This distinguishes the action coefficient from
the combinatorial vertex normalisation. The reaction diagrams show
incoming and outgoing excitations; the displaced-field vertices also
include the loss and response contributions required by the master
equation. The resulting action specifies the effective field theory.
A determination of its scaling regime or universality class would
require a separate analysis of these vertices and their renormalisation.
\section{Pomeron-driven odderon evolution and nonlinear saturation}
\label{sec:meanfield_analysis}

In this section we use the mean-field Eqs. (\ref{eq:sec2:meanP}) and (\ref{eq:sec2:meanO}) to analyse the pomeron catalytic amplification of the odderon, given a non zero odd initial seed; the odderon saturation due to its  ternary fusion.

The pomeron equation is autonomous,
whereas the odderon evolves on the pomeron background. This triangular
structure follows from the selected reactions; it does not require a
small-feedback approximation. The complex moments need not represent
positive populations. Real positive profiles provide a useful regime
in which the amplification and saturation mechanisms can be displayed
explicitly. We first set $D_P=D_O=0$, so that evolution is local in
impact parameter.

\subsection{The logistic pomeron background}
The pomeron field obeys
\begin{equation}
 \partial_\tau P=aP-rP^2.
 \label{eq:analysis:P_background}
\end{equation}
For constant $a\ne0$ and $r\ne0$, its analytic solution with initial profile
$P_{\rm in}(b)$ at $\tau_0$ is
\begin{align}
P(\tau,b)&=\frac{P_{\rm in}(b)e^{a\Delta\tau}}{B(\tau,b)},
 \qquad \Delta\tau=\tau-\tau_0,\nonumber\\
 B(\tau,b)&=1+\frac{rP_{\rm in}(b)}a   \left(e^{a\Delta\tau}-1\right).
 \label{eq:analysis:P_solution}
\end{align}
This is the logistic evolution used in the preceding construction
\cite{Kohara:2026}. For real positive $a,r$ and strictly positive
$P_{\rm in}(b)$, it tends to $P_{\rm sat}=a/r$ at each fixed $b$.
Complex initial profiles are evolved by the same expression on intervals
where $B$ remains nonzero. The identity
\begin{equation}
 \partial_\tau\log B(\tau,b)=rP(\tau,b)
 \label{eq:analysis:B_identity}
\end{equation}
will also determine the accumulated catalytic amplification.

\subsection{Analytic odderon evolution with ternary fusion}
On this background the odderon equation becomes
\begin{equation}
 \partial_\tau O(\tau,b)=2kP(\tau,b)O(\tau,b)-2uO(\tau,b)^3.
 \label{eq:analysis:O_nonlinear}
\end{equation}
The catalytic process increases the number of odderons by two per event,
while ternary fusion reduces it by two. Neither process produces a
nonzero field from $O_{\rm in}=0$. The model thus describes the evolution
of a seed component rather than a source of an odd field from a purely
even initial condition.

Let's define
\begin{equation}
A(\tau,b)=2k\int_{\tau_0}^{\tau}P(\tau',b)\,d\tau'
 =\frac{2k}{r}\operatorname{Log}B(\tau,b).
 \label{eq:analysis:O_integral}
\end{equation}
For nonzero $O$, we introduce $Y=O^{-2}$. The direct differentiation gives
\begin{eqnarray}
&& \partial_\tau Y
=-2O^{-3}\partial_\tau O=-4kPY+4u,
\nonumber\\
&&
\partial_\tau\big(e^{2A}Y\big)=4u e^{2A}.
\label{eq:analysis:O_inverse}
\end{eqnarray}
Integrating from $\tau_0$ then yields \begin{equation}
 O(\tau,b)=
 \frac{O_{\rm in}(b)e^{A(\tau,b)}}{\sqrt{1+4uO_{\rm in}(b)^2  \displaystyle\int_{\tau_0}^{\tau}e^{2A(\tau',b)}\,d\tau'}}.
 \label{eq:analysis:O_solution}
\end{equation}
The branch of the square root is fixed by continuity from its value
$1$ at $\tau_0$. This expression extends continuously to a zero seed,
for which $O$ remains zero. Using the logistic background gives
\begin{equation}
 O(\tau,b)=
 \frac{O_{\rm in}(b)B(\tau,b)^{2k/r}}
 {\sqrt{1+4uO_{\rm in}(b)^2
       \displaystyle\int_{\tau_0}^{\tau}B(\tau',b)^{4k/r}\,d\tau'}}.
 \label{eq:analysis:O_logistic_solution}
\end{equation}
This is an analytic solution in quadrature. The integral is evaluated
numerically for general parameters. For complex profiles,
$B^\nu=\exp[\nu\operatorname{Log}B]$ is continued from $B(\tau_0,b)=1$.
The solution is considered on intervals avoiding zeros of $B$ and of
the square-root argument. The resulting dependence is odd in
$O_{\rm in}$, with the same continuation used for opposite seeds.

When the integrated fusion correction is small,
\begin{equation}
 \left|4uO_{\rm in}(b)^2  \int_{\tau_0}^{\tau}e^{2A(\tau',b)}\,d\tau'\right|\ll1,
 \label{eq:analysis:linear_interval}
\end{equation}
Eq.~\eqref{eq:analysis:O_solution} reduces to
\begin{equation}
 O(\tau,b)\simeq O_{\rm in}(b)
 \exp\!\left[2k\int_{\tau_0}^{\tau}P(\tau',b)\,d\tau'\right]
 =O_{\rm in}(b)B(\tau,b)^{2k/r} ~.
 \label{eq:analysis:O_linear:solution}
\end{equation}
The corresponding instantaneous balance is determined by the ratio
$|uO^2|/|kP|$. Amplification is initially catalytic when this ratio is
small, whereas ternary fusion becomes essential as it approaches unity.

\subsection{Saturation and a subdominant odderon component}
For a constant positive background $P=\overline P$ and real positive
$k,u$, the stationary nonzero fields satisfy
\begin{equation}
 O_{\rm eq}^2=\frac{k\overline P}{u}.
 \label{eq:analysis:instantaneous_plateau}
\end{equation}
The exact solution in this constant background is
\begin{equation}
O(\tau)=\frac{O_{\rm in}e^{2k\overline P\Delta\tau}}
{\sqrt{1+\dfrac{uO_{\rm in}^2}{k\overline P}    \left(e^{4k\overline P\Delta\tau}-1\right)}}.
 \label{eq:analysis:constant_background}
\end{equation}
A positive initial field approaches $+\sqrt{k\overline P/u}$ and a
negative one approaches the opposite value. These statements refer to
real initial fields; they are not a general bound on complex solutions.

On the growing logistic background, the local threshold
$\sqrt{kP(\tau,b)/u}$ varies with rapidity. Below it a positive odderon
field grows, whereas above it the field decreases. For strictly positive
initial profiles and real positive coefficients, the asymptotic limits
at each fixed $b$ are
\begin{equation}
 P\longrightarrow P_{\rm sat}=\frac ar,
 \qquad
 O\longrightarrow O_{\rm sat}=\sqrt{\frac{ka}{ur}}.
 \label{eq:analysis:saturation}
\end{equation}
The asymptotic ratio is therefore
\begin{equation}
 \frac{O_{\rm sat}}{P_{\rm sat}}
 =\sqrt{\frac{kr}{ua}}.
 \label{eq:analysis:small_component}
\end{equation}
A small stationary component is obtained when $kr/(ua)\ll1$.
This condition controls the plateau ratio, while finite-rapidity ratios also
depend on the initial profiles. Linearization around either nonzero
stationary branch gives relaxation eigenvalues $-a$ and $-4ka/r$.
Thus both branches are locally stable in the real homogeneous system.

\subsection{Evolution of transverse profiles}
To give the reader a qualitative representation of our approach, we create initial profiles with a decreasing tail in impact parameter. We test the profiles typically used in the modeling of elastic scattering. 
We compare Gaussian and regularized exponential initial shapes,
\begin{align}
 P_{\rm in}(b)&=P_c F(b),\qquad O_{\rm in}(b)=O_c F(b),\nonumber\\
 F_G(b)&=e^{-b^2/R_{\rm rms}^2},\label{eq:analysis:initial_gaussian}\\
 F_E(b)&=\exp\!\left[-\frac{\sqrt{b^2+b_c^2}-b_c}{L}\right].
 \label{eq:analysis:initial_profiles}
\end{align}
As showed in our previous work \cite{Kohara:2026} under these profiles our equation satisfies the Froissart bound \cite{Froissart:1961}.
The exponential profile is regular at the origin and has an exponential
large-$b$ tail. The decay length $L$ is chosen to match the Gaussian
radial mean square radius with measure $d^2b=2\pi b\,db$:
\begin{equation}
 \langle b^2\rangle_E
 =\frac{6L^3+6b_cL^2+2b_c^2L}{L+b_c}=R_{\rm rms}^2.
 \label{eq:analysis:matched_radii}
\end{equation}
Both species have the same initial shape within each family; the two
families share central values and radial mean square radius.

\begin{figure}[h]
 %\centering
\includegraphics[width=\linewidth]{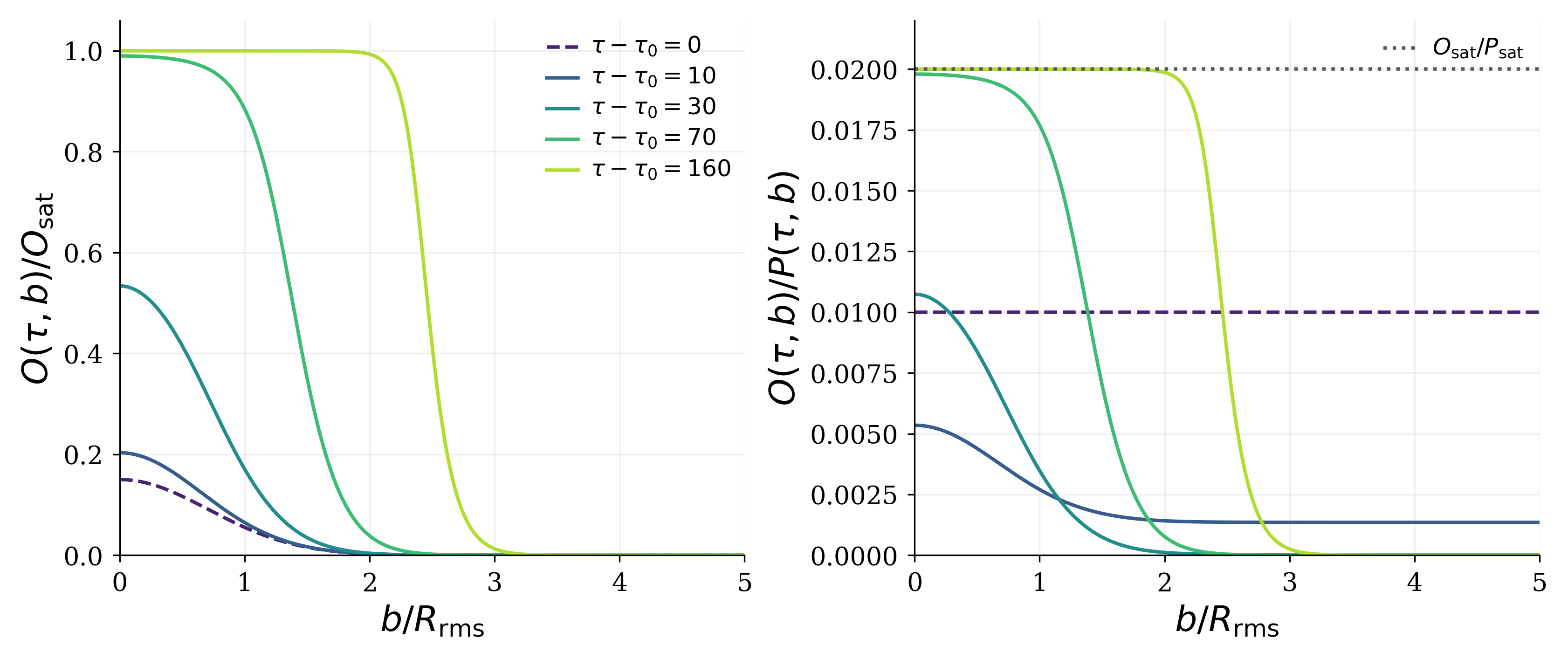}
 \caption{Catalytic amplification with ternary fusion for Gaussian initial
 profiles. Left: $O(\tau,b)$ divided by $O_{\rm sat}=\sqrt{ka/(ur)}$.
 Right: the local ratio $O/P$. The dotted horizontal line marks the
 asymptotic ratio $\sqrt{kr/(ua)}=0.02$. The dashed curve is the initial
 profile. All fields and coefficients are real and positive.}
 \label{fig:analysis:gaussian}
\end{figure}

\begin{figure}[h]
 \centering
 \includegraphics[width=\linewidth]{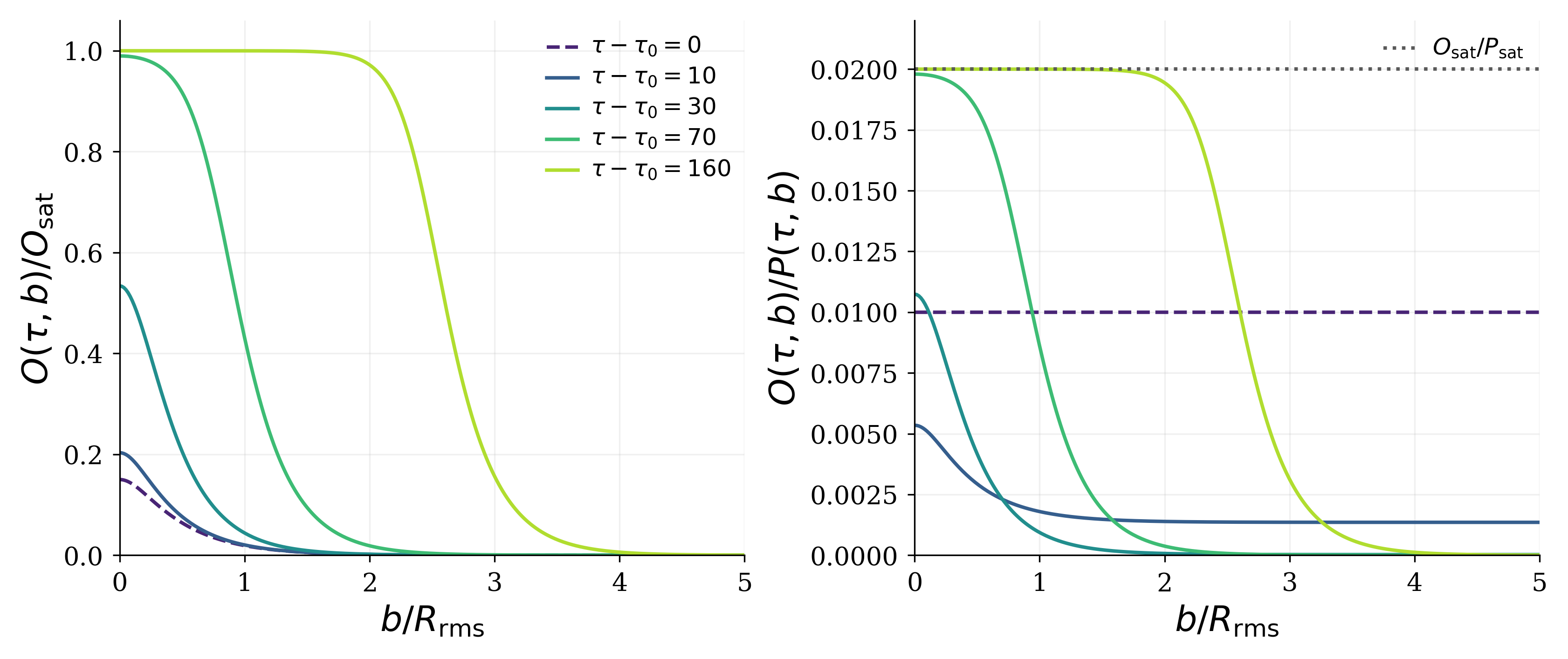}
 \caption{Same as Fig.~\ref{fig:analysis:gaussian}, with regularised
 exponential initial profiles of the same central values and radial mean
 square radius.}
 \label{fig:analysis:exponential}
\end{figure}

The figures use scaled, dimensionless variables with
$a=r=0.2$, $k=0.03$, $u=75$, $P_c=0.3$, $O_c=0.003$,
$R_{\rm rms}=1$ and $b_c=0.2$. These parameters give
$P_{\rm sat}=1$, $O_{\rm sat}=0.02$ and
$O_{\rm sat}/P_{\rm sat}=0.02$; they serve to illustrate the dynamics.
Figures~\ref{fig:analysis:gaussian} and~\ref{fig:analysis:exponential}
show the profiles at $\Delta\tau=0,10,30,70,160$ and the corresponding
local ratio $O/P$. For these parameters the ratio stays below $2\%$
over the displayed domain.

\begin{figure}[h]
 \centering
 \includegraphics[width=\linewidth]{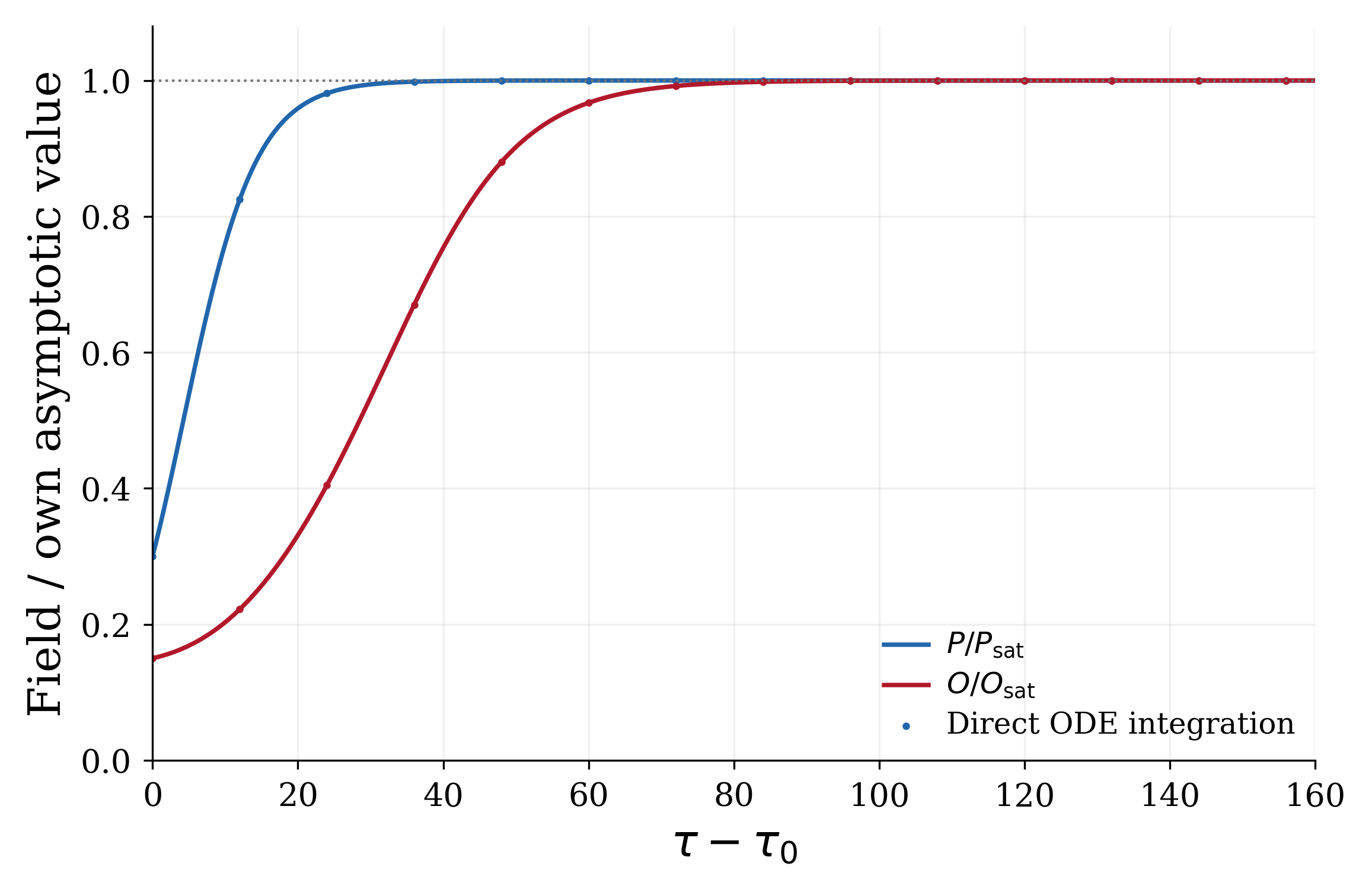}
 \caption{Central evolution towards the pomeron and odderon plateaus.
 Each field is divided by its own asymptotic value; the equal plateau
 heights in the plot therefore correspond to $O_{\rm sat}/P_{\rm sat}=0.02$.
 Solid curves show the analytic solutions evaluated by quadrature;
 points show an independent direct integration of the same mean-field
 equations.}
 \label{fig:analysis:central}
\end{figure}

The central region saturates first. At larger impact parameters the
smaller seed fields delay amplification and the formation of a plateau.
This outward development occurs even without diffusion because the
initial profiles have nonzero tails. Pointwise saturation at fixed $b$
is not uniform over the transverse plane and does not imply saturation
of its integral. Figure~\ref{fig:analysis:central} shows the central
rapidity evolution. The two shape families give identical central curves
because they share central initial conditions and the evolution is local.
Independent integration of Eq.~\eqref{eq:sec2:meanO} with zero
diffusion checks the analytic solution in quadrature.


\section{From correlation functions to elastic-scattering amplitudes}
\label{sec:amplitude_matching}

The complex moments provide evolving transverse profiles. To connect
them to elastic scattering, we specify an effective matching to even and
odd exchange amplitudes. The covariance of the closed equations permits
opposite odderon seeds to evolve into opposite profiles on the same
pomeron background. This property supplies a consistent mean-field
basis for the matching used below. The matching fixes the physical
interpretation of the profiles; it is additional to their definition as
weighted configuration moments.

\subsection{Complex moments and the exchange convention}
The first moments are
\begin{equation}
 G_A(b,\tau)=\langle s|\hat p_A(b)|\Phi(\tau)\rangle,
 \qquad A=P,O.
 \label{eq:matching:first_moments}
\end{equation}
Before the continuum cell-area rescaling they read
\begin{equation}
 g_{A,i}(\tau)=\sum_{\boldsymbol n}n_{A,i} \mathcal W(\boldsymbol n;\tau).
 \label{eq:matching:weighted_moments}
\end{equation}
For positive normalised weights they are mean occupations. With complex
weights they are complex first factorial moments, with no required
positive-density interpretation. This distinction applies to both
species and separates the general framework from the real positive
examples displayed above.

At the level of the closed effective equations we adopt the following amplitudes correspondences
\begin{equation}
 T_+(\tau,b)=i\mathcal N_P P(\tau,b),
 \qquad
 T_-(\tau,b)=-\mathcal N_O O(\tau,b),
 \label{eq:matching:first_moment_prescription}
\end{equation}
where $\mathcal N_P$ and $\mathcal N_O$ are real positive constants,
including the dimensions needed to obtain dimensionless amplitude
profiles. The external-channel matching assigns the same initial $P$
profile and opposite initial $O$ profiles to conjugate hadronic channels.
Equation~\eqref{eq:sec2:meanO} 
preserves this assignment, giving
an even $T_+$ and an odd $T_-$ under this channel transformation.
These identifications are effective matching assumptions, not operator
identities equating an occupation moment to an exact hadronic amplitude.

The factors in Eq.~\eqref{eq:matching:first_moment_prescription} implement
the desired relative phase convention. Writing $G_A=X_A+iY_A$ gives
\begin{align}
 \operatorname{Re}T_+&=-\mathcal N_PY_P,&
 \operatorname{Im}T_+&=\mathcal N_PX_P,\nonumber\\
 \operatorname{Re}T_-&=-\mathcal N_OX_O,&
 \operatorname{Im}T_-&=-\mathcal N_OY_O.
\label{eq:matching:phases}
\end{align}
Predominantly real moments therefore describe an imaginary-leading even
exchange and a real-leading odd exchange. The pomeron equation becomes
\begin{equation}
 \partial_\tau T_+=D_P\nabla_b^2T_++aT_++i\frac r{\mathcal N_P}T_+^2,
\label{eq:matching:pomeron_evolution}
\end{equation}
recovering the preceding effective evolution equation
\cite{Kohara:2026}. For the odd profile the same matching gives
\begin{equation}
 \partial_\tau T_-=D_O\nabla_b^2T_-
 -\frac{2ik}{\mathcal N_P}T_+T_-
 -\frac{2u}{\mathcal N_O^2}T_-^3.
 \label{eq:matching:odderon_evolution}
\end{equation}
Every term is odd under $T_-\mapsto-T_-$ at fixed $T_+$.
The real stationary result of the previous section corresponds to a
predominantly real odd amplitude in this convention. General complex
solutions retain their phases and the analyticity restrictions stated
above. It is important to recall that the real-field plateau does not establish a general complex bound
or impose impact-parameter unitarity by itself.

The microscopic number-parity conservation and the effective transformation
of the closed fields are distinct statements. In particular, the covariance
of the first-moment closure does not prove the same simple transformation
for all exact weighted moments. Beyond this matching approximation,
hadronic boundary matrix elements can involve higher correlators, as in
stochastic Reggeon constructions \cite{Kolevatov:2011}.
QCD definitions of odd exchange likewise specify charge-odd scattering
operators \cite{Hatta:2005}. The present prescription
provides the mean-field profiles used for the following sensitivity
analysis, rather than an independent derivation of hadronic impact factors.

\subsection{Sensitivity of the forward ratio}
For azimuthally symmetric profiles, the even and odd scattering amplitudes
are related to their impact-parameter representations by a common
Fourier--Bessel convention,
\begin{equation}
 F_\pm(s,t)=\mathcal K(s)\,2\pi\int_0^\infty b\,db\,
 J_0(b\sqrt{-t})\,T_\pm(\tau,b),\qquad t\leq0,
 \label{eq:matching:transform}
\end{equation}
where $\tau-\tau_0=\log(s/s_0)$ and $\mathcal K(s)$ is a common positive
normalisation factor. The two physical channels are
\begin{equation}
 F_{pp}=F_++F_-,\qquad F_{p\bar p}=F_+-F_- ~,
 \label{eq:matching:channels}
\end{equation}
where $F_{pp}$ represents the proton-proton scattering amplitude and $F_{p \bar p}$  the proton-antiproton one. 
The complex phases of the evolved moments and their matching to
$T_\pm$ determine the interference between the two exchange sectors.

At $t=0$, we write $F_\pm=R_\pm+iI_\pm$ and we define
$\rho_+=R_+/I_+$, with $I_+\ne0$. The exact channel shifts are
\begin{align}
 \rho_{pp}-\rho_+&=\frac{R_--\rho_+I_-}{I_++I_-},\label{eq:matching:rho_exact_pp}\\
 \rho_{p\bar p}-\rho_+&=-\frac{R_--\rho_+I_-}{I_+-I_-}.
 \label{eq:matching:rho_exact}
\end{align}
For an odd amplitude small compared with $|I_+|$, these expressions give
\begin{equation}
 \rho_{pp,p\bar p}-\rho_+
 \simeq\pm\frac{R_--\rho_+I_-}{I_+}.
 \label{eq:matching:rho_linear}
\end{equation}
A predominantly real odd exchange therefore shifts the two forward ratios
in opposite directions at leading order. Its contribution can be visible
in $\rho$ even when it is small relative to the dominant imaginary even
amplitude. The sign of this shift depends on the initial moments and the
external matching. In particular, the reaction network alone does not
impose an ordering between $\rho_{pp}$ and $\rho_{p\bar p}$.
The transverse integrals and the relative normalisation of the two
exchange sectors enter this comparison. A local ratio of moments does
not by itself determine the ratio of forward amplitudes.

\subsubsection{An illustrative energy evolution}
\label{sec:rho_numerical_illustration}
To examine this sensitivity, consider the retained reaction network
without transverse diffusion. In terms of $Y=\log(s/s_0)$, its local
mean-field equations are
\begin{equation}
 \partial_Y P=aP-rP^2,\qquad
 \partial_Y O=2kPO-2uO^3.
 \label{eq:rho:local_system}
\end{equation}
Here $P$ and $O$ denote scaled complex first moments. We use the
dimensionless radial coordinate $x=b/R$, where $R$ is a common transverse
scale, and take $\sqrt{s_0}=10$ GeV. The initial conditions are modeled 
\begin{align}
 P_{\rm in}(x)&=P_c e^{-x^2}[1-iq_0(x)],\nonumber\\
 q_0(x)&=q_c+\frac{q_s}{1+\exp[-(x-x_c)/w]},\nonumber\\
 O_{\rm in}(x)&=O_c e^{-(x/R_O)^2},
 \label{eq:rho:profiles}
\end{align}
with the dimensionless parameters
\begin{align}
 a=r&=0.79,& k&=0.03,&u&=0.02,\nonumber\\
 P_c&=0.0046,&O_c&=0.00031,&R_O&=0.7,\nonumber\\
 q_c&=0.005,&q_s&=0.62,&x_c&=1.5,&w&=0.1.
 \label{eq:rho:parameters}
\end{align}
The parameters were chosen to approximately reproduce selected reference values of the forward ratio and illustrate its nonmonotonic energy dependence and the splitting between the \(pp\) and \(p\bar p\) channels, without performing a global statistical fit.
The magnitude of the initial pomeron phase increases towards the
periphery, while the real odderon seed is concentrated more closely
around the origin. %The parameters specify an example near selected forward-ratio reference values; the resulting curves illustrate the dynamics rather than provide an independent fit or prediction of those values.

\begin{figure}[tbp]
% \centering
\includegraphics[width=1.1\linewidth]{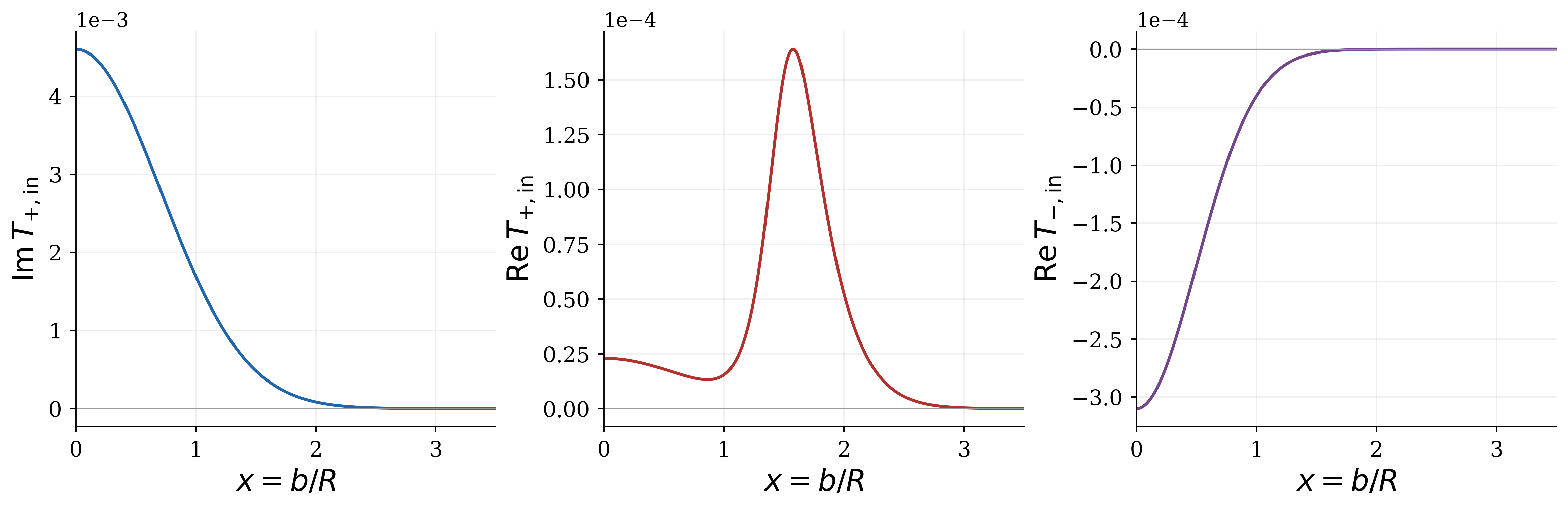}
 \caption{Initial impact-parameter amplitude profiles at
 $\sqrt{s_0}=10$ GeV. From left to right:
 $\operatorname{Im}T_{+,\rm in}=\operatorname{Re}P_{\rm in}$,
 $\operatorname{Re}T_{+,\rm in}=-\operatorname{Im}P_{\rm in}$,
 and $\operatorname{Re}T_{-,\rm in}=-O_{\rm in}$.
 The initial imaginary odd component vanishes. The panels use separate
 vertical scales and show the unrescaled profile values.}
 \label{fig:rho:initial_profiles}
\end{figure}

For this example, the matching is
\begin{equation}
 T_+=iP,\qquad T_-=-O.
 \label{eq:rho:matching}
\end{equation}
The fields retain their identification with the scaled first
moments. The minus sign belongs to the amplitude matching,
$T_-=-O$, and does not redefine the odderon correlator.
Since its evolution equation is odd in $O$, reversing the
initial odderon seed reverses the evolved odd amplitude while
leaving the pomeron solution unchanged.  Figure~\ref{fig:rho:initial_profiles}
shows the initial amplitude profiles associated with
Eq.~\eqref{eq:rho:profiles}. The Gaussian $\operatorname{Re}P_{\rm in}$
is the positive imaginary part of $T_{+,\rm in}$, whereas
$-\operatorname{Im}P_{\rm in}$ supplies its smaller real part.
The initial odd amplitude is real and negative in this convention.

The local evolution has an analytic solution in quadrature. With
\begin{equation}
 B(Y,x)=1+\frac{rP_{\rm in}(x)}a(e^{aY}-1),\qquad
 I(Y,x)=\int_0^Y B(y,x)^{4k/r}\,dy,
 \label{eq:rho:quadrature}
\end{equation}
the two fields are
\begin{equation}
 P(Y,x)=\frac{P_{\rm in}(x)e^{aY}}{B(Y,x)},\qquad
 O(Y,x)=\frac{O_{\rm in}(x)B(Y,x)^{2k/r}}
 {\sqrt{1+4uO_{\rm in}(x)^2I(Y,x)}}.
 \label{eq:rho:solutions}
\end{equation}
The powers and square root are continued from their initial values.
For the profiles and rapidities considered here, their arguments remain
in the right half-plane, so this continuation is implemented by the
principal branches. The complex pomeron background generates an
imaginary odderon component during evolution even though its seed is real.

Define the transverse integrals
\begin{equation}
 \mathcal P(Y)=\int_0^\infty x\,dx\,P(Y,x),\qquad
 \mathcal O(Y)=\int_0^\infty x\,dx\,O(Y,x).
\label{eq:rho:integrated_fields}
\end{equation}
We recall that the Bessel function $J_0(0)=1$.
The common factor $\mathcal K(s)2\pi R^2$ cancels in the forward ratios,
which become
\begin{align}
 \rho_{pp}&=\frac{-\operatorname{Im}\mathcal P-\operatorname{Re}\mathcal O}{\operatorname{Re}\mathcal P-       \operatorname{Im}\mathcal O},\nonumber\\
 \rho_{p\bar p}&=\frac{-\operatorname{Im}\mathcal P+       \operatorname{Re}\mathcal O}      {\operatorname{Re}\mathcal P+       \operatorname{Im}\mathcal O},\qquad
 \rho_+=-\frac{\operatorname{Im}\mathcal P}    {\operatorname{Re}\mathcal P}.
\label{eq:rho:ratios}
\end{align}
Thus the fields are integrated over the transverse plane before the
ratios are formed.

The rise and subsequent decrease of $\rho_+$ originate from the radial
phase structure. Writing $P=X\,\left[1-i\,q(Y,b)\right]$, with $X>0$. For real positive $a$ and
$r$, the local logistic equation gives
\begin{equation}
 \partial_Y q=-rXq(1+q^2).
 \label{eq:rho:local_phase}
\end{equation}
Each positive local phase ratio therefore decreases. In contrast,
\begin{equation}
\rho_+=\int x\,dx\,Xq/\int x\,dx\,X 
\end{equation}
 is a transverse average with
evolving weights. The centre saturates before the periphery, whose
larger initial phase ratio acquires an increased relative weight.
For the present initial conditions, this redistribution first raises
the integrated ratio. At higher energies, the local phase relaxation
dominates and the forward ratio decreases.

\begin{figure}[tbp]
 \centering
 \includegraphics[width=\linewidth]{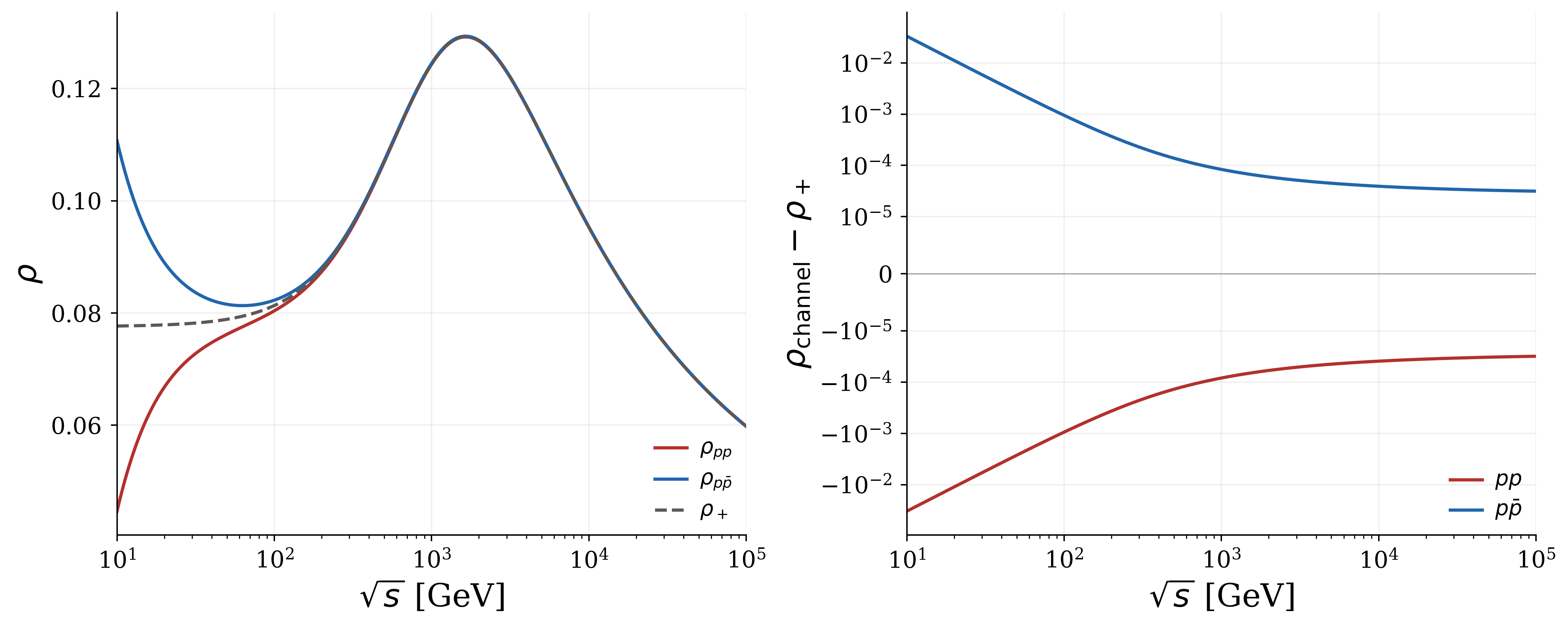}
 \caption{Forward ratios for the original Gaussian seeds and the complex
 pomeron phase of Eq.~\eqref{eq:rho:profiles}. Left: $\rho_{pp}$ (red),
 $\rho_{p\bar p}$ (blue), and the even-sector ratio $\rho_+$ (dashed grey).
 Right: their odd-exchange shifts relative to $\rho_+$; the ordinate is
 symmetric logarithmic with a linear threshold of $10^{-5}$.
 These curves use a fixed illustrative parameter set and are not compared
 with experimental data.}
 \label{fig:rho:calibration}
 \label{fig:rho:correction}
\end{figure}

Figure~\ref{fig:rho:calibration} displays this evolution and the odd
channel splitting. Representative values are
\begin{equation}
\begin{array}{c|ccc}
 \sqrt{s}\ [\mathrm{GeV}]&\rho_+&\rho_{pp}&\rho_{p\bar p}\\\hline
 52&0.07895&0.07643&0.08147\\
 1800&0.12912&0.12906&0.12918\\
 13000&0.08969&0.08965&0.08972
\end{array}
 \label{eq:rho:anchor_values}
\end{equation}
The $pp$ ratio reaches its maximum near $1.65$ TeV. The odd contribution
lowers $\rho_{pp}$ and raises $\rho_{p\bar p}$, with shifts of
approximately $\mp0.0025$ at $52$ GeV and
$\mp3.8\times10^{-5}$ at $13$ TeV. The high-energy decrease in this
example is consequently driven mainly by the even sector, while the
odderon produces the channel splitting.

The finite-energy regime sampled here remains far from odderon
saturation: $u|O|^2/(k|P|)$ stays below $1.4\times10^{-5}$.
The example therefore primarily probes phase evolution and catalytic
amplification. It does not determine the fusion coefficient $u$ or
establish an asymptotically small odd component. This distinction also
matters for scattering unitarity. Under the direct matching
\eqref{eq:rho:matching} and the convention $S=1+iT$, the two channel
conditions are
\begin{equation}
 G_{\rm inel}^{pp,p\bar p}
 =2\operatorname{Im}(iP\mp O)-|iP\mp O|^2\geq0.
 \label{eq:rho:unitarity}
\end{equation}
The inequalities are checked on the finite energy and radial grids
used for the illustration. They constrain the scattering
interpretation independently of conservation of the total
Doi--Peliti weight.

For the real homogeneous stationary solution and the direct
matching used here, the inelastic overlap is
\begin{equation}
 G_{\mathrm{inel},*}
 =2P_*-P_*^2-O_*^2
 =P_*\left(2-P_*-\frac{k}{u}\right),
 \qquad P_*=\frac{a}{r}.
\end{equation}
For positive couplings and $0<P_*<2$, its non-negativity requires
\begin{equation}
 u\geq\frac{k}{2-a/r}.
\end{equation}
This is a necessary stationary condition, not a sufficient
criterion for unitarity throughout the complex evolution.

The illustrative parameters give $P_*=1$ and $k/u=1.5$, hence
$G_{\mathrm{inel},*}=-0.5$. The example is therefore restricted
to the finite-energy regime examined here and cannot be
extrapolated into a globally unitary asymptotic model with
the same parameters and matching. Establishing such an extension
requires conditions on both the couplings and the complex
initial profiles.

\subsection{Sensitivity near a diffractive minimum}
The forward ratio and the diffractive region depend on different
transverse projections of the evolved fields. At $t=0$, the Bessel
kernel in Eq.~\eqref{eq:matching:transform} equals unity. At nonzero
transfer it can produce cancellations between radial regions and reduce
the even amplitude. An odd exchange that is small in the forward
direction can then become relatively more important near a minimum.

The exact decomposition
\begin{equation}
 |F_+\pm F_-|^2=|F_+|^2
 \pm2\operatorname{Re}(F_+F_-^*)+|F_-|^2
 \label{eq:matching:dip}
\end{equation}
shows that the effect is controlled by both the interference and the
odd quadratic term. The latter must be retained near a sufficiently
small even amplitude. A real odd contribution can change the residual
real amplitude where the imaginary even component crosses zero, so the
minimum may become shallower, deepen, or shift. The two channels differ by
\begin{equation}
 |F_{pp}|^2-|F_{p\bar p}|^2
 =4\operatorname{Re}(F_+F_-^*).
 \label{eq:matching:channel_difference}
\end{equation}
For exactly imaginary $F_+$ and real $F_-$ the interference vanishes,
and the odd quadratic contribution changes both channels equally.
Likewise, at an exact even-amplitude zero both channels have intensity
$|F_-|^2$, although their neighbouring profiles can differ.

The same initial conditions used for $\rho$ allow this sensitivity to be
examined without introducing additional transverse structure. We define
$z=|t|R^2$ and the reduced amplitudes
\begin{align}
 \mathfrak F_+(Y,z)&=i\int_0^\infty x\,dx\,
 J_0(x\sqrt z)P(Y,x),\label{eq:matching:reduced_transform:even}\\
 \mathfrak F_-(Y,z)&=-\int_0^\infty x\,dx\,
 J_0(x\sqrt z)O(Y,x).
\label{eq:matching:reduced_transform}
\end{align}
Then $F_\pm=\mathcal K(s)2\pi R^2\mathfrak F_\pm$. We plot
\begin{equation}
 \mathcal D_{pp,p\bar p}(Y,z)
 =|\mathfrak F_+(Y,z)\pm\mathfrak F_-(Y,z)|^2.
 \label{eq:matching:reduced_intensity}
\end{equation}
These intensities are proportional to the differential cross sections
through a common, unspecified dimensional factor. They are not divided
by the forward even-sector intensity. Since $R$ is not fixed, the
horizontal axis remains $|t|R^2$ rather than a physical transfer scale
in $\mathrm{GeV}^2$.

\begin{figure}[h]
 \centering
 \includegraphics[width=\linewidth]{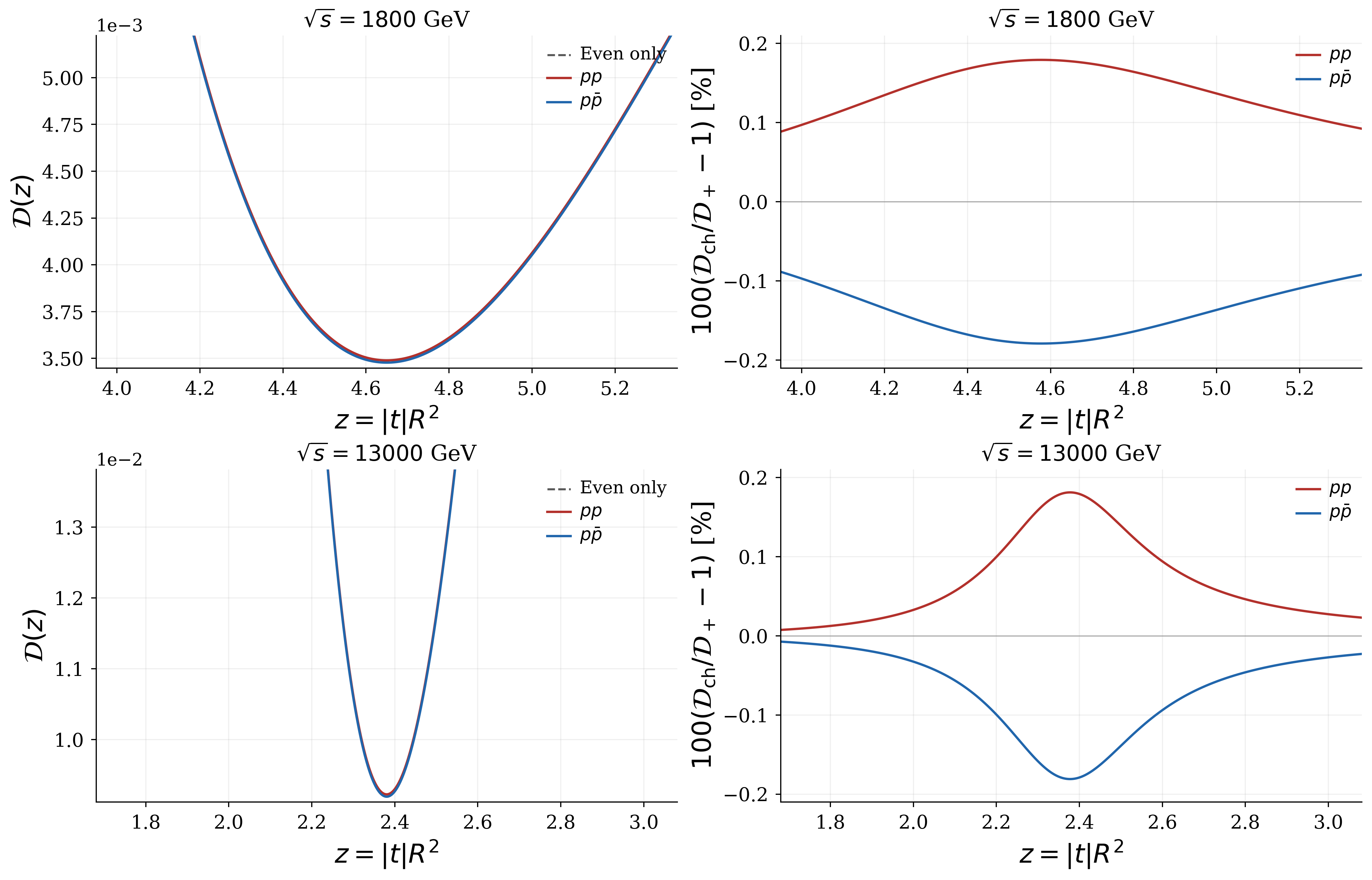}
 \caption{First diffractive minima for the same original profiles used
 in Fig.~\ref{fig:rho:calibration}. The upper and lower rows correspond
 to $\sqrt{s}=1.8$ and $13$ TeV. Left: reduced intensities
 $\mathcal D_{pp}$ (red), $\mathcal D_{p\bar p}$ (blue), and
 $\mathcal D_+=|\mathfrak F_+|^2$ (dashed grey).
 Right: relative corrections to the even-sector intensity, shown to
 resolve the small channel splitting. The calculation uses neither
 an enhanced external odd coupling nor a modified seed profile.}
 \label{fig:matching:dip_original}
\end{figure}

Figure~\ref{fig:matching:dip_original} shows the first minima at
$\sqrt{s}=1.8$ and $13$ TeV. The even-sector minima occur near
$z=4.650$ and $2.381$, respectively. Evaluated at those positions, the
odd contribution changes the $pp$ intensities by approximately
$+0.177\%$ and $+0.181\%$, with nearly opposite changes in $p\bar p$.
The chosen phase therefore gives a slightly shallower $pp$ minimum in
this example. Neither the magnitude nor the ordering is universal.

The weak separation demonstrates that the forward-ratio illustration
does not by itself provide a quantitative description of the
channel-dependent dip structure. Measurements of both channels around
$53$ GeV provide a direct experimental reference in this region
\cite{Breakstone:1985}; at TeV energies, the D0--TOTEM comparison uses
an extrapolation of the $pp$ data to the $p\bar p$ measurement energy
\cite{Abazov:2021}. Constraining complex initial amplitudes from such
information would require a dedicated analysis of their transverse
shapes and phases. Here the forward and diffractive observables identify
complementary consequences of the matched evolution: the former
illustrates a channel splitting in the real-to-imaginary ratio, while
the latter exposes its dependence on the relative phase and on radial
cancellations beyond the forward limit.

%\clearpage

\section{Conclusions and perspectives}
\label{sec:conclusions}

We have constructed an effective reaction-based description of coupled
pomeron and odderon evolution in rapidity and impact parameter using the
Doi--Peliti formalism. Pomeron branching and merging generate the logistic
even-sector equation, while pomeron-catalysed odderon branching and ternary
odderon fusion provide amplification and, in the real positive mean-field
regime, nonlinear regulation of the odd sector. The master equations, their operator representation and the
coherent-state functional integral describe the same evolution. The
hierarchy of factorial moments makes explicit the correlations discarded
by the mean-field closure.

Allowing complex configuration weights extends the algebraic description
beyond positive probabilities while retaining conservation of their total
weight. The mean-field equations are covariant under $O\mapsto-O$ with
$P$ fixed. In the reduced network, the pomeron evolves autonomously and
controls the amplification and phase evolution of an initially nonzero
odderon seed. The network does not generate a nonzero first odd moment
from an identically vanishing odd field at mean-field level.

The matching to scattering amplitudes connects the evolved moments to
$\rho$ and to interference near a diffractive minimum. With simple
Gaussian initial profiles, the example exhibits a separation between
$\rho_{pp}$ and $\rho_{p\bar p}$ and a rise followed by a decrease of the
forward ratio. The changing transverse weights and pomeron phase account
for the nonmonotonic behaviour, while the odd amplitude produces the
channel splitting. The dip correction remains small and depends on the
relative phase and transverse profiles. These results illustrate a chosen
initial-value problem rather than establish a quantitative description
of the measured channel difference.

Soft elastic scattering remains a fundamental probe of strong interactions
in a regime where a quantitative calculation directly from QCD is
difficult, despite substantial phenomenological progress and
QCD-inspired non-perturbative descriptions
\cite{Block:2006,Dosch:1994}. Its experimental study is also demanding:
measurements of the very forward scattering require access to the Coulomb--nuclear interference
region, with control of beam optics, alignment, normalisation and the
hadronic amplitude used in the extraction
\cite{Antchev:2019,Aad:2023}. Extracting $\rho$, however,
requires assumptions about the hadronic amplitude, which introduce
some model dependence \cite{Kohara:2017ats,Kohara:2019
,Kohara:2022ktp}. Measurements over a wider
momentum-transfer range provide complementary information on the
diffractive structure \cite{Antchev:2019b}. The limitation is therefore
not an absence of experiments, but sparse coverage in energy and a lack
of matched high-energy $pp$ and $p\bar p$ measurements.

The renewed interest in the odderon illustrates the importance of this
limitation. The ISR measurements provided a direct comparison of the
two channels near $53$ GeV \cite{Breakstone:1985}, whereas the D0--TOTEM
comparison at TeV energies uses an extrapolation of the $pp$ measurements
to the $p\bar p$ energy \cite{Abazov:2021}. Independent extrapolation
analyses have also assessed the evidence \cite{Cui:2023}. These studies
motivate further work on how energy evolution and its uncertainties
enter the comparison. Once calibrated and tested at independent energies,
the present equation could provide a complementary dynamical prescription
for transporting amplitude profiles, with uncertainties from the initial
conditions, couplings and mean-field approximation explicitly retained.

More generally, separating the initial profiles from their evolution
offers a way to compare phenomenological models under a common dynamical
law. Different admissible profiles could be tested for which features
persist, which are washed out and which lead to distinct predictions
at unmeasured energies. Such comparisons would require a common fitting
strategy and uncertainty propagation; they would test the proposed
evolution rather than assume its universality. This perspective also
connects Regge dynamics to nonlinear systems and to developments in
high-energy QCD. The established relations between RFT and stochastic
processes \cite{Cardy:1980}, between nonlinear QCD evolution and
travelling waves \cite{Munier:2003}, and between QCD evolution and
Reggeon field theories \cite{Kovner:2007} provide useful starting points.
They motivate comparisons with the present construction without
constituting a QCD derivation of its reaction network.

Further theoretical work must identify domains preserved by the full
complex evolution and compatible with scattering unitarity, and assess
corrections beyond mean field. Experimentally, additional energies,
overlapping momentum-transfer coverage, and published covariance
information would sharpen tests of both the even and odd sectors.
Continued measurements from the interference region through the dip
would therefore constrain the evolution itself, as well as the profiles
at individual energies. Elastic scattering remains an opportunity to
test non-perturbative strong-interaction dynamics through a small set
of precise, complementary observables.

\appendix

\section{Derivation of the master equations}
\label{AppendixA}

\subsection{Lattice conventions and configuration weights}
We first work on a transverse lattice. At one cell, write
$m=n_P$, $n=n_O$ and $W_{m,n}(\tau)$ for the configuration weight,
with all spectator occupations held fixed. Weights with negative
occupation labels vanish. The unnormalised Fock basis satisfies
\begin{align}
 |m,n\rangle&=\hat q_P^m\hat q_O^n|0\rangle,\nonumber\\
 \hat q_P|m,n\rangle&=|m+1,n\rangle,&
 \hat p_P|m,n\rangle&=m|m-1,n\rangle,\nonumber\\
 \hat q_O|m,n\rangle&=|m,n+1\rangle,&
 \hat p_O|m,n\rangle&=n|m,n-1\rangle.
 \label{eq:A:basis}
\end{align}
The commutators are $[\hat p_A,\hat q_B]=\delta_{AB}$ at a single
cell, with an additional $\delta_{ij}$ on the full lattice. The state
and projection bra are
\begin{equation}
 |\Phi\rangle=\sum_{\boldsymbol n}\mathcal W(\boldsymbol n;\tau)
 |\boldsymbol n\rangle,
 \qquad
 \langle s|=\langle0|\exp\!\left(\sum_{A,i}\hat p_{A,i}\right).
 \label{EQ:Appendix_Fock_sum}
\end{equation}
All changes of gain-term indices below are performed inside complete
sums. The equations are algebraic and apply to complex configuration
weights, provided the relevant sums exist.

The retained local reactions and their lattice propensities are
\begin{equation}
\begin{array}{c|c|c}
 \text{reaction}&\text{coefficient}&\text{propensity at }(m,n)\\ \hline
 P\to2P&a_\ell&a_\ell m\\
 2P\to P&r_\ell&r_\ell m(m-1)\\
 P+O\to P+3O&k_\ell&k_\ell mn\\
 3O\to O&u_\ell&u_\ell n(n-1)(n-2)
\end{array}
 \label{eq:A:channels}
\end{equation}
The subscript $\ell$ distinguishes lattice coefficients from continuum
coefficients. Factors $1/2!$ or $1/3!$ for unordered incoming sets,
if used, are absorbed in $r_\ell$ or $u_\ell$. We use
$\partial_\tau|\Phi\rangle=-\hat H|\Phi\rangle$ throughout.

\subsection{Pomeron branching: $P\to2P$}
An event increases $m$ by one. The gain at $(m,n)$ comes from $(m-1,n)$,
where $m-1$ pomerons can branch. Thus
\begin{equation}
 \left.\partial_\tau W_{m,n}\right|_a
 =a_\ell\left[(m-1)W_{m-1,n}-mW_{m,n}\right].
 \label{EQ:Master_Equation_Splitting}
\end{equation}
Multiplication by the basis ket and summation give
\begin{align}
 \left.\partial_\tau|\Phi\rangle\right|_a
 &=a_\ell\sum_{m,n\geq0}
 \left[(m-1)W_{m-1,n}|m,n\rangle-mW_{m,n}|m,n\rangle\right]
 \nonumber\\
 &=a_\ell\sum_{m,n\geq0}mW_{m,n}
\left(|m+1,n\rangle-|m,n\rangle\right).
 \label{EQ:Master_Equation_Splitting_Fock_2}
\end{align}
The second line uses $m-1\mapsto m$ in the gain contribution.
The required operator actions are
\begin{equation}
 \hat q_P^2\hat p_P|m,n\rangle=m|m+1,n\rangle,
 \qquad
 \hat q_P\hat p_P|m,n\rangle=m|m,n\rangle.
\end{equation}
Consequently,
\begin{equation}
 \left.\partial_\tau|\Phi\rangle\right|_a
 =a_\ell(\hat q_P^2-\hat q_P)\hat p_P|\Phi\rangle,
 \qquad
 \hat H_a=a_\ell(\hat q_P-\hat q_P^2)\hat p_P.
\label{EQ:Master_Equation_Splitting_Fock_3}
\end{equation}

\subsection{Pomeron merging: $2P\to P$}
An event reduces $m$ by one. The source configuration $(m+1,n)$
has propensity $r_\ell(m+1)m$, giving
\begin{equation}
 \left.\partial_\tau W_{m,n}\right|_r
 =r_\ell\left[(m+1)mW_{m+1,n}-m(m-1)W_{m,n}\right].
 \label{EQ:Master_Equation_Merging}
\end{equation}
Inside the configuration sum, shifting $m+1\mapsto m$ yields
\begin{align}
\left.\partial_\tau|\Phi\rangle\right|_r
 &=r_\ell\sum_{m,n\geq0}
 \left[(m+1)mW_{m+1,n}|m,n\rangle-m(m-1)W_{m,n}|m,n\rangle\right]\nonumber\\
 &=r_\ell\sum_{m,n\geq0}m(m-1)W_{m,n}\left(|m-1,n\rangle-|m,n\rangle\right).
 \label{EQ:Master_Equation_Merging_Fock_2}
\end{align}
Since
\begin{align}
 \hat q_P\hat p_P^2|m,n\rangle&=m(m-1)|m-1,n\rangle,\nonumber\\
 \hat q_P^2\hat p_P^2|m,n\rangle&=m(m-1)|m,n\rangle,
\end{align}
we obtain
\begin{equation}
 \left.\partial_\tau|\Phi\rangle\right|_r
 =r_\ell(\hat q_P-\hat q_P^2)\hat p_P^2|\Phi\rangle,
 \qquad
 \hat H_r=r_\ell(\hat q_P^2-\hat q_P)\hat p_P^2.
 \label{EQ:Master_Equation_Merging_Fock_3}
\end{equation}

\subsection{Pomeron-catalysed odderon branching: $P+O\to P+3O$}
This reaction leaves $m$ unchanged and raises $n$ by two. The master
equation is
\begin{equation}
 \left.\partial_\tau W_{m,n}\right|_k
 =k_\ell m\left[(n-2)W_{m,n-2}-nW_{m,n}\right].
 \label{EQ:Master_Equation_PO_to_P3O}
\end{equation}
Negative source labels give no gain, and the gain also vanishes at $n=2$
because its source has no odderon. Multiplying and summing explicitly,
\begin{align}
 \left.\partial_\tau|\Phi\rangle\right|_k
 &=k_\ell\sum_{m,n\geq0}m\left[
 (n-2)W_{m,n-2}|m,n\rangle-nW_{m,n}|m,n\rangle\right]
 \nonumber\\
 &=k_\ell\sum_{m,n\geq0}mnW_{m,n}
\left(|m,n+2\rangle-|m,n\rangle\right).
 \label{eq:A:catalytic_shift}
\end{align}
The gain-term shift is $n-2\mapsto n$. The operator actions are
\begin{align}
 \hat q_P\hat q_O^3\hat p_P\hat p_O|m,n\rangle
 &=mn|m,n+2\rangle,\nonumber\\
 \hat q_P\hat q_O\hat p_P\hat p_O|m,n\rangle
 &=mn|m,n\rangle.
\end{align}
The creation operator $\hat q_P$ restores the pomeron removed by
$\hat p_P$, while $\hat p_P$ supplies the catalytic factor $m$.
Therefore
\begin{align}
 \left.\partial_\tau|\Phi\rangle\right|_k
 &=k_\ell\hat q_P(\hat q_O^3-\hat q_O)\hat p_P\hat p_O|\Phi\rangle,\nonumber\\
 \hat H_k&=k_\ell\hat q_P(\hat q_O-\hat q_O^3)\hat p_P\hat p_O.
 \label{EQ:Fock_PO_to_P3O}
\end{align}

\subsection{Ternary odderon fusion: $3O\to O$}
The reaction lowers $n$ by two. The gain source $(m,n+2)$ has propensity
$u_\ell(n+2)(n+1)n$, so
\begin{equation}
 \left.\partial_\tau W_{m,n}\right|_u
 =u_\ell\left[(n+2)(n+1)nW_{m,n+2}
-n(n-1)(n-2)W_{m,n}\right].
 \label{EQ:Master_Equation_3O_to_O}
\end{equation}
The summed contribution and its shifted form are
\begin{align} \left.\partial_\tau|\Phi\rangle\right|_u
&=u_\ell\sum_{m,n\geq0}\left[
 (n+2)(n+1)nW_{m,n+2}|m,n\rangle\right.\nonumber\\
 &\hspace{38mm}\left.-n(n-1)(n-2)W_{m,n}|m,n\rangle\right]
 \nonumber\\
&=u_\ell\sum_{m,n\geq0}n(n-1)(n-2)W_{m,n}               \left(|m,n-2\rangle-|m,n\rangle\right).
\label{eq:A:ternary_shift}
\end{align}
The gain shift is $n+2\mapsto n$. The two operator actions are
\begin{align}
 \hat q_O\hat p_O^3|m,n\rangle
 &=n(n-1)(n-2)|m,n-2\rangle,\nonumber\\
 \hat q_O^3\hat p_O^3|m,n\rangle
 &=n(n-1)(n-2)|m,n\rangle.
\end{align}
Thus
\begin{align}
 \left.\partial_\tau|\Phi\rangle\right|_u
 &=u_\ell(\hat q_O-\hat q_O^3)\hat p_O^3|\Phi\rangle,\nonumber\\
 \hat H_u&=u_\ell(\hat q_O^3-\hat q_O)\hat p_O^3.
 \label{EQ:Fock_3O_to_O}
\end{align}
The factor $n(n-1)(n-2)$ sets the loss to zero for $n<3$.
In particular, this reaction does not remove an isolated odderon.

\subsection{Diffusion and the continuum coefficients}
For a particle of species $A$ hopping from $j$ to a neighbouring cell $i$
with rate $h_{A,ij}$ per particle, the full-configuration master equation is
\begin{equation}
 \left.\partial_\tau\mathcal W(\boldsymbol n)\right|_{j\to i}
 =h_{A,ij}\left[(n_{A,j}+1)
 \mathcal W(\boldsymbol n+\boldsymbol e_{A,j}-\boldsymbol e_{A,i})-n_{A,j}\mathcal W(\boldsymbol n)\right].
 \label{EQ:Master_Equation_Diffusion_cell}
\end{equation}
At $n_{A,i}=0$ the source of the gain term has a negative label and
its weight vanishes. Changing the summation variable to the source
configuration $\boldsymbol m$ gives
\begin{align}
 \left.\partial_\tau|\Phi\rangle\right|_{j\to i}
 &=h_{A,ij}\sum_{\boldsymbol m}m_{A,j}\mathcal W(\boldsymbol m)
 \left(|\boldsymbol m-\boldsymbol e_{A,j}+\boldsymbol e_{A,i}\rangle
-|\boldsymbol m\rangle\right)\nonumber\\&=h_{A,ij}(\hat q_{A,i}-\hat q_{A,j})\hat p_{A,j}|\Phi\rangle.
 \label{EQ:Master_Equation_Diffusion}
\end{align}
The pair $(i,j)$ is ordered, so reverse hops are included separately.
For a square lattice of spacing $a_\perp$, take
$h_{A,ij}=D_A/a_\perp^2$ for each ordered nearest-neighbour hop. Then
\begin{align}
 \hat H_{\rm diff}
 &=\sum_{A,(i,j)}\frac{D_A}{a_\perp^2}(\hat q_{A,j}-\hat q_{A,i})\hat p_{A,j},\nonumber\\
\left.\partial_\tau\langle\hat p_{A,i}\rangle\right|_{\rm diff}
 &=\frac{D_A}{a_\perp^2}\sum_{j\sim i}
       \left(\langle\hat p_{A,j}\rangle-\langle\hat p_{A,i}\rangle\right).
 \label{EQ:diffusion_continuum}
\end{align}

The field and rate rescalings are defined in
Eqs.~\eqref{eq:hier:field_scaling} and~\eqref{eq:hier:rate_scaling}.
With cell area $v=a_\perp^2$, they give the formal continuum hopping
Hamiltonian $-\sum_A D_A\int d^2b\,\hat q_A\nabla_b^2\hat p_A$.
The reaction polynomials above carry the corresponding continuum
coefficients after the same rescaling. Local products retain the
lattice prescription, and spatial surface terms are eliminated by
the chosen boundary conditions.

\subsection{Total generator and conservation of the weight}
At one cell the reaction generator is
\begin{align}
 \hat L_{\rm reac}={}&a_\ell(\hat q_P^2-\hat q_P)\hat p_P
 +r_\ell(\hat q_P-\hat q_P^2)\hat p_P^2\nonumber\\
 &+k_\ell\hat q_P(\hat q_O^3-\hat q_O)\hat p_P\hat p_O
 +u_\ell(\hat q_O-\hat q_O^3)\hat p_O^3.
 \label{EQ:Master-Equation_Sum}
\end{align}
The full generator is obtained by summing over cells and ordered hops;
$\hat H=-\hat L$. Since $\langle s|\hat q_{A,i}=\langle s|$,
each creation-operator difference in the reaction and hopping terms
vanishes upon left projection. Therefore
\begin{equation}
 \langle s|\hat H=0,
 \qquad
 \partial_\tau\sum_{\boldsymbol n}\mathcal W(\boldsymbol n;\tau)
 =-\langle s|\hat H|\Phi\rangle=0.
 \label{EQ:normalisation_appendix}
\end{equation}
This property holds separately for every channel and remains valid for
complex weights. An initial total weight equal to one stays equal to one;
its rapidity derivative is zero.

\section{Commutators, factorial moments and symmetry}
\label{Appendix_B}

\subsection{Evolution identity and factorial observables}
For a time-independent operator $\hat F$, define
$\langle\hat F\rangle=\langle s|\hat F|\Phi\rangle$. Conservation of
the projection bra gives
\begin{equation}
 \partial_\tau\langle\hat F\rangle
 =-\langle s|\hat F\hat H|\Phi\rangle
 =\langle s|[\hat H,\hat F]|\Phi\rangle.
 \label{eq:B:commutator}
\end{equation}
For a diagonal configuration observable $F(\boldsymbol n)$, the equivalent
increment formula is
\begin{equation}
\left.\partial_\tau\langle F\rangle\right|_\rho
 =\left\langle R_\rho(\boldsymbol n)
       [F(\boldsymbol n+\boldsymbol\nu_\rho)-F(\boldsymbol n)]\right\rangle,
 \label{eq:B:increments}
\end{equation}
where $R_\rho$ is the reaction propensity and $\boldsymbol\nu_\rho$
its occupation change. For annihilation-operator moments the associated
configuration observable is a falling factorial:
\begin{equation}
 M_{h,j}=\langle\hat p_P^h\hat p_O^j\rangle =\langle(m)_h(n)_j\rangle,
 \qquad (n)_j=n(n-1)\cdots(n-j+1),\quad(n)_0=1.
 \label{eq:B:factorial_definition}
\end{equation}
No positive-weight assumption is required for these identities.

\subsection{Explicit first-moment commutators}
For a normal-ordered polynomial,
$[F(\hat q_P,\hat q_O),\hat p_A]=-\partial_{\hat q_A}F$.
The local reaction Hamiltonian from Appendix~\ref{AppendixA} gives
\begin{align}
 [\hat H_a,\hat p_P]&=a_\ell(2\hat q_P-1)\hat p_P,&
 [\hat H_a,\hat p_O]&=0,\nonumber\\
 [\hat H_r,\hat p_P]&=r_\ell(1-2\hat q_P)\hat p_P^2,&
 [\hat H_r,\hat p_O]&=0,\nonumber\\
 [\hat H_k,\hat p_P]&=k_\ell(\hat q_O^3-\hat q_O)\hat p_P\hat p_O,&
 [\hat H_k,\hat p_O]&=k_\ell\hat q_P(3\hat q_O^2-1)\hat p_P\hat p_O,
 \nonumber\\
 [\hat H_u,\hat p_P]&=0,&
 [\hat H_u,\hat p_O]&=u_\ell(1-3\hat q_O^2)\hat p_O^3.
 \label{eq:B:explicit_commutators}
\end{align}
Projection replaces the normally ordered creation operators by one.
The catalytic contribution to the pomeron first moment thus vanishes,
whereas its contribution to the odderon first moment is
$2k_\ell\langle\hat p_P\hat p_O\rangle$. We obtain
\begin{align}
 \partial_\tau\langle\hat p_P\rangle
 &=a_\ell\langle\hat p_P\rangle-r_\ell\langle\hat p_P^2\rangle,
 \label{eq:B:GP}\\
 \partial_\tau\langle\hat p_O\rangle
 &=2k_\ell\langle\hat p_P\hat p_O\rangle
-2u_\ell\langle\hat p_O^3\rangle.
 \label{eq:B:GO}
\end{align}
These equations can also be checked directly from
Eq.~\eqref{eq:B:increments}: the changes in $(m,n)$ are $(1,0)$,
$(-1,0)$, $(0,2)$ and $(0,-2)$ for the four channels.

For an ordered hop,
\begin{equation}
 [\hat H_{j\to i},\hat p_{A,l}]
 =h_{A,ij}(\delta_{il}-\delta_{jl})\hat p_{A,j}.
 \label{eq:B:diff_commutator}
\end{equation}
Summing gives the discrete Laplacian. After continuum rescaling,
\begin{align}
 \partial_\tau G_P(b)&=D_P\nabla_b^2G_P(b)
+aG_P(b)-rG_{PP}(b,b),\nonumber\\
 \partial_\tau G_O(b)&=D_O\nabla_b^2G_O(b)
+2kG_{PO}(b,b)-2uG_{OOO}(b,b,b).
 \label{eq:B:continuum_exact}
\end{align}
Rapidity arguments are implicit. There is no remaining Dirac delta
multiplying these equations: the delta from a commutator has already
been integrated over the Hamiltonian coordinate.

These exact equations yield
Eqs.~\eqref{EQ:Evolution_Equation_Correlator_Pomeron} and
\eqref{EQ:Evolution_Equation_Correlator_Odderon} in the main text.
The factorisation and closed mean-field equations are introduced only
in Section~\ref{sec:hierarchy}. The pomeron marginal is autonomous
before closure because both odderon reactions leave its occupations
unchanged; its exact factorial hierarchy still requires closure.

\subsection{Higher factorial moments}
The exact local hierarchy follows either from commutators or from the
increment identity. For a reaction with incoming vector
$X=(X_P,X_O)$ and outgoing vector $Y=(Y_P,Y_O)$, the propensity convention
of Appendix~\ref{AppendixA} gives
\begin{equation}
 \left.\partial_\tau M_{h,j}\right|_{X\to Y}
 =\lambda_\ell
 \sum_{\substack{0\leq s\leq h,\ 0\leq t\leq j\\s+t\geq1}}
 \binom hs\binom jt
 \left[(Y_P)_s(Y_O)_t-(X_P)_s(X_O)_t\right]
 M_{h-s+X_P,\,j-t+X_O}.
 \label{eq:B:general_hierarchy}
\end{equation}
Terms with zero coefficient are omitted before interpreting moment
indices. A direct derivation uses
\begin{equation}
 \partial_\tau Z=\lambda_\ell
 (x^{Y_P}y^{Y_O}-x^{X_P}y^{X_O})
 \partial_x^{X_P}\partial_y^{X_O}Z
\end{equation}
and differentiates $h$ times in $x$ and $j$ times in $y$ at $x=y=1$.
Leibniz' rule gives Eq.~\eqref{eq:B:general_hierarchy}.

For the retained network this specialises to
\begin{align}
 \left.\partial_\tau M_{h,j}\right|_{\rm reac}
 ={}&a_\ell\left[hM_{h,j}+h(h-1)M_{h-1,j}\right]\nonumber\\
 &-r_\ell\left[hM_{h+1,j}+h(h-1)M_{h,j}\right]\nonumber\\
&+k_\ell\sum_{t=1}^{\min(3,j)}\binom jt d_t\left[M_{h+1,j-t+1}+hM_{h,j-t+1}\right]\nonumber\\
 &-u_\ell\left[2jM_{h,j+2}+3j(j-1)M_{h,j+1}
+j(j-1)(j-2)M_{h,j}\right],
\label{eq:B:network_hierarchy}
\end{align}
where $d_1=2$, $d_2=6$, $d_3=6$. A term multiplied by a zero prefactor
is absent. This hierarchy is exact at one cell without diffusion.

For example, $(m)_2=m(m-1)$ changes under pomeron branching by
$(m+1)_2-(m)_2=2m$, while merging gives
$(m-1)_2-(m)_2=-2(m-1)$. Multiplication by the respective propensities
and use of $m^2=(m)_2+m$ and
$m(m-1)^2=(m)_3+(m)_2$ give
\begin{equation}
 \partial_\tau M_{2,0}
 =2a_\ell(M_{2,0}+M_{1,0})
  -2r_\ell(M_{3,0}+M_{2,0}).
 \label{eq:B:P2}
\end{equation}
For the odderon second factorial moment,
\begin{equation}
 (n+2)_2-(n)_2=4n+2,
 \qquad (n-2)_2-(n)_2=-4n+6.
\end{equation}
Using $mn(4n+2)=4m(n)_2+6mn$ and
$(n)_3(-4n+6)=-4(n)_4-6(n)_3$ yields
\begin{equation}
 \partial_\tau M_{0,2}
 =4k_\ell M_{1,2}+6k_\ell M_{1,1}
  -4u_\ell M_{0,4}-6u_\ell M_{0,3}.
 \label{eq:B:O2}
\end{equation}
Finally, the mixed first-by-first moment obeys
\begin{equation}
 \partial_\tau M_{1,1}
 =a_\ell M_{1,1}-r_\ell M_{2,1}
+2k_\ell(M_{2,1}+M_{1,1})-2u_\ell M_{1,3}.
 \label{eq:B:PO}
\end{equation}
Thus the closed first-moment equations are not the exact hierarchy.
In particular, zero first odderon moment for a general complex ensemble
need not imply zero mixed and higher odd-factorial moments. A microscopically
empty odderon sector, with support only at $n_O=0$, stays empty, whereas
the corresponding statement for a zero field in the mean-field equations
refers to the chosen closure.

On a multi-cell lattice, define
$M_{h,j}^{(i)}=\langle\hat p_{P,i}^h\hat p_{O,i}^j\rangle$.
For symmetric nearest-neighbour hopping the extra terms are
\begin{align}
 \left.\partial_\tau M_{h,j}^{(i)}\right|_{\rm diff}
 ={}&\frac{hD_P}{a_\perp^2}\sum_{l\sim i}
 \left[\langle\hat p_{P,i}^{h-1}\hat p_{P,l}\hat p_{O,i}^j\rangle
-M_{h,j}^{(i)}\right]\nonumber\\
 &+\frac{jD_O}{a_\perp^2}\sum_{l\sim i}
 \left[\langle\hat p_{P,i}^h\hat p_{O,i}^{j-1}\hat p_{O,l}\rangle
-M_{h,j}^{(i)}\right].
 \label{eq:B:diffusion_hierarchy}
\end{align}
A line is omitted if its prefactor $h$ or $j$ is zero. Diffusion therefore
introduces separated-site correlators, not just a Laplacian of a local
higher moment. The continuum expressions are taken after rescaling
moments and rates with the cell area.

\subsection{Number parity and covariance of the closure}
The one-cell generating function and its reaction evolution are
\begin{align}
 Z(x,y;\tau)&=\sum_{m,n\geq0}W_{m,n}(\tau)x^my^n,\nonumber\\
 \partial_\tau Z=\mathcal L Z,\qquad
 \mathcal L={}&a_\ell(x^2-x)\partial_x+r_\ell(x-x^2)\partial_x^2\nonumber\\
 &+k_\ell x(y^3-y)\partial_x\partial_y
       +u_\ell(y-y^3)\partial_y^3.
 \label{eq:B:generatorZ}
\end{align}
Introduce $\mathsf C Z(x,y)=Z(x,-y)$. Under this operation, $y$ and
$\partial_y$ each change sign. The two odderon polynomials are odd,
as are the corresponding derivative factors. Every term in
$\mathcal L$ is therefore unchanged and
\begin{equation}
 [\mathcal L,\mathsf C]=0,
 \qquad Z_\pm(x,y)=\frac{Z(x,y)\pm Z(x,-y)}2.
 \label{eq:B:number_parity}
\end{equation}
The parity sectors evolve independently. Diffusion conserves the total
odderon number and hence the global number parity on the lattice;
it need not preserve the parity at an individual cell.

The factorial moments are derivatives at the projection point:
\begin{equation}
 M_{h,j}=\left.\partial_x^h\partial_y^j Z(x,y)\right|_{x=y=1}.
 \label{eq:B:moments_from_Z}
\end{equation}
For example, differentiating the parity-transformed function gives
\begin{equation}
 \left.\partial_y Z(x,-y)\right|_{x=y=1}
 =-\left.\partial_y Z(x,y)\right|_{x=1,y=-1},
 \label{eq:B:projection_point}
\end{equation}
which is generally different from $-M_{0,1}$. Thus the conserved
microscopic number parity does not alone impose the transformation
$G_O\mapsto-G_O$ on a first moment taken with the standard projection bra.

The additional field covariance of the mean-field closure is stated
in Eq.~\eqref{eq:hier:field_covariance}. It is distinct from the exact
number-parity relation derived here. No commutation between factorial
factorisation and number-parity projection is assumed.

\section*{Acknowledgments}

I am deeply grateful to Josephine M. Kohara for
insightful discussions and support. I warmly  thank Pedro C. Malta for reading the manuscript and suggesting significant improvements.
I thank OpenAI's ChatGPT (model GPT-6.1 Sol) for assistance with refining the English phrasing and language flow in selected sections of the manuscript, as well as the assistance in generating Python code for the illustrative figures. All outputs were checked and revised by author, who takes full responsibility for the content of this work.

%\bibliography{references}

\begin{thebibliography}{53}


\bibitem{Kohara:2026}
A.~K.~Kohara,
``Complex logistic equation for universal energy evolution in hadronic elastic scattering,''
Phys.\ Rev.\ D \textbf{113} (2026) 116041,
doi:10.1103/xz23-bqg9.


\bibitem{BSW:1979} C. Bourrely, J. Soffer, and Tai Tsun Wu. ``New impact picture for low- and high-energy proton-proton elastic scattering.'' Phys.\ Rev.\ D, 19:3249--3260, 1979.


\bibitem{Islam:2009}
M. M. Islam, V. Innocente, T. Fearnley and G. Sanguinetti,
 ``High-Energy $\bar p p$ and $pp$ Elastic Scattering and Nucleon Structure'',
 Europhys.\ Lett.\ \textbf{4} (1987) 189--196, doi:10.1209/0295-5075/4/2/010.




\bibitem{kohara:2014} A. K. Kohara, E. Ferreira, and T. Kodama. ``pp Elastic Scattering at LHC Energies.'' Eur. Phys. J. C, 74:3175, 2014,
doi:10.1140/epjc/s10052-014-3175-x.

\bibitem{Kohara:2019qoq} A. K. Kohara. ``Forward scattering amplitudes of $pp$ and $\bar{p}p$ with crossing symmetry and scaling properties.'' J. Phys. G, 46(12):125001, 2019,
doi:10.1088/1361-6471/ab47d3.

\bibitem{Csorgo:2013usx} T. Csorgo and F. Nemes. ``Elastic scattering of protons from $\sqrt{s} = 23.5$ GeV to 7 TeV from a generalized Bialas-Bzdak model.''  Int. J. Mod. Phys. A, 29:1450019, 2014.



\bibitem{Selyugin:2023txo} O. V. Selyugin. ``New properties of elastic $pp$ and $\bar{p}p$ scattering at high energies.'' \textit{Eur. Phys. J. C}, 84(6):649, 2024,
doi:10.1140/epjc/s10052-024-13017-9.

\bibitem{Donnachie:2013xia} A. Donnachie and P. V. Landshoff. ``$pp$ and $\bar{p}p$ total cross sections and elastic scattering.'' Phys. Lett. B, 727:500--505, 2013,
doi:10.1016/j.physletb.2013.10.068.


\bibitem{Collins:1977}
P.~D.~B.~Collins,
\textit{An Introduction to Regge Theory and High Energy Physics},
Cambridge University Press, Cambridge (1977).

\bibitem{Abarbanel:1975}
H.~D.~I.~Abarbanel, J.~B.~Bronzan, R.~L.~Sugar and A.~R.~White,
``Reggeon field theory: Formulation and use,''
Phys.\ Rept.\ \textbf{21} (1975) 119--182,
doi:10.1016/0370-1573(75)90034-4.

\bibitem{Regge:1959}
T.~Regge,
``Introduction to complex orbital momenta,''
Nuovo Cimento \textbf{14} (1959) 951--976,
doi:10.1007/BF02728177.


\bibitem{Chew:1961}
G.~F.~Chew and S.~C.~Frautschi,
``Principle of Equivalence for All Strongly Interacting Particles Within the S-Matrix Framework,''
Phys.\ Rev.\ Lett.\ \textbf{7} (1961) 394--397,
doi:10.1103/PhysRevLett.7.394.

\bibitem{Gribov:1961}
V.~N.~Gribov,
``Partial waves with complex orbital angular momenta and the asymptotic
behavior of the scattering amplitude,''
Sov.\ Phys.\ JETP \textbf{14} (1962) 1395--1400,
[Zh.\ Eksp.\ Teor.\ Fiz.\ \textbf{41} (1961) 1962--1970].



\bibitem{Gribov:1967}
V.~N.~Gribov,
``A Reggeon diagram technique,''
Zh.\ Eksp.\ Teor.\ Fiz.\ \textbf{53} (1967) 654--672,
[Sov.\ Phys.\ JETP \textbf{26} (1968) 414--423].



\bibitem{Antchev:2019}
G.~Antchev \textit{et al.} [TOTEM Collaboration],
``First determination of the $\rho$ parameter at $\sqrt{s}=13$ TeV:
probing the existence of a colourless C-odd three-gluon compound state,''
Eur.\ Phys.\ J.\ C \textbf{79} (2019) 785,
doi:10.1140/epjc/s10052-019-7223-4,
arXiv:1812.04732 [hep-ex].

\bibitem{Abazov:2021}
V.~M.~Abazov \textit{et al.} [D0 and TOTEM Collaborations],
``odderon Exchange from Elastic Scattering Differences between $pp$ and
$p\bar p$ Data at 1.96 TeV and from $pp$ Forward Scattering Measurements,''
Phys.\ Rev.\ Lett.\ \textbf{127} (2021) 062003,
doi:10.1103/PhysRevLett.127.062003.

\bibitem{Kwiecinski:1980}J. Kwiecinski and  M. Praszalowicz, ``Three gluon integral equation and odd C singlet Regge singularities in QCD.'' Physics \ Lett. \ B, \textbf{94} (1980) 413-416.

\bibitem{Bartels:1980}J. Bartels, ``High-energy behavior in a non-Abelian gauge theory (II).'' Nucl. \ Phys. \ B, \textbf{175} (1980) 365-401.

\bibitem{Ewerz:2003xi}
C.~Ewerz,
``The odderon in quantum chromodynamics,''
arXiv:hep-ph/0306137.



\bibitem{Kovchegov:2004}
Y.~V.~Kovchegov, L.~Szymanowski and S.~Wallon,
``Perturbative Odderon in the dipole model,''
Phys.\ Lett.\ B \textbf{586} (2004) 267--281,
doi:10.1016/j.physletb.2004.02.036,
arXiv:hep-ph/0309281.

\bibitem{Hatta:2005}
Y.~Hatta, E.~Iancu, K.~Itakura and L.~McLerran,
``Odderon in the color glass condensate,''
Nucl.\ Phys.\ A \textbf{760} (2005) 172--207,
doi:10.1016/j.nuclphysa.2005.05.163,
arXiv:hep-ph/0501171.

\bibitem{Kakkad:2022}
H.~Kakkad, A.~K.~Kohara and P.~Kotko,
``Evolution equation for elastic scattering of hadrons,''
Eur.\ Phys.\ J.\ C \textbf{82} (2022) 830,
doi:10.1140/epjc/s10052-022-10747-6.


\bibitem{Grassberger:1978}
P.~Grassberger and K.~Sundermeyer,
``Reggeon field theory and Markov processes,''
Phys.\ Lett.\ B \textbf{77} (1978) 220--222,
doi:10.1016/0370-2693(78)90626-3.

\bibitem{Cardy:1980}
J.~L.~Cardy and R.~L.~Sugar,
``Directed percolation and Reggeon field theory,''
J.\ Phys.\ A \textbf{13} (1980) L423--L427,
doi:10.1088/0305-4470/13/12/002.

\bibitem{Bondarenko:2007}
S.~Bondarenko, L.~Motyka, A.~H.~Mueller, A.~I.~Shoshi and B.-W.~Xiao,
``Exploiting the equivalence of reggeon field theory in zero transverse
dimensions and reaction--diffusion processes,''
Eur.\ Phys.\ J.\ C \textbf{50} (2007) 593--601,
doi:10.1140/epjc/s10052-007-0218-6,
arXiv:hep-ph/0609213.

\bibitem{Kolevatov:2011}
R.~S.~Kolevatov, K.~G.~Boreskov and L.~V.~Bravina,
``All-loop calculation of the Reggeon field theory amplitudes via stochastic model,''
Eur.\ Phys.\ J.\ C \textbf{71} (2011) 1757,
doi:10.1140/epjc/s10052-011-1757-4,
arXiv:1105.3673 [hep-ph].

\bibitem{Bartels:2017}
J.~Bartels, C.~Contreras and G.~P.~Vacca,
``Pomeron--Odderon interactions in a reggeon field theory,''
Phys.\ Rev.\ D \textbf{95} (2017) 014013,
doi:10.1103/PhysRevD.95.014013,
arXiv:1608.08836 [hep-th].

\bibitem{Braun:2021}
M.~A.~Braun, E.~M.~Kuzminskii and M.~I.~Vyazovsky,
``Local one-dimensional reggeon model of the interaction of pomerons and odderons,''
Eur.\ Phys.\ J.\ C \textbf{81} (2021) 676,
doi:10.1140/epjc/s10052-021-09467-0.

\bibitem{Braun:2025}
M.~A.~Braun,
``Probabilities in Toy Regge models with odderons,''
Eur.\ Phys.\ J.\ C \textbf{85} (2025) 1400,
doi:10.1140/epjc/s10052-025-15142-5.

\bibitem{Doi:1976} 
M. Doi, Second quantization representation for classical many-particle system, J. Phys. A 9, 1465–1477 (1976). 


\bibitem{Peliti:1985}
L. Peliti, Path integral approach to birth-death processes on a lattice, J. Physique 46, 1469–1483 (1985).



\bibitem{Tauber:2005}
U.~C.~T\"auber, M.~Howard and B.~P.~Vollmayr-Lee,
``Applications of field-theoretic renormalization group methods to reaction--diffusion problems,''
J.\ Phys.\ A \textbf{38} (2005) R79--R131,
doi:10.1088/0305-4470/38/17/R01,
arXiv:cond-mat/0501678.

\bibitem{Weber:2017}
M.~F.~Weber and E.~Frey,
``Master equations and the theory of stochastic path integrals,''
Rept.\ Prog.\ Phys.\ \textbf{80} (2017) 046601,
doi:10.1088/1361-6633/aa5ae2,
arXiv:1609.02849 [cond-mat.stat-mech].

\bibitem{Alessandrini:1977} V. Alessandrini, D. Amati, and M. Ciafaloni. ``Classical Kinks and their Quantization in Supercritical Reggeon Field Theory. '' \textit{Nucl. Phys. B}, 130:429--485, 1977.

\bibitem{Balitsky:1996}
I.~Balitsky,
``Operator expansion for high-energy scattering,''
Nucl.\ Phys.\ B \textbf{463} (1996) 99,
doi:10.1016/0550-3213(95)00638-9,
arXiv:hep-ph/9509348.

\bibitem{Weigert:2005}
H.~Weigert,
``Evolution at small $x_{\mathrm{bj}}$:
The Color Glass Condensate,''
Prog.\ Part.\ Nucl.\ Phys.\ \textbf{55} (2005) 461--565,
doi:10.1016/j.ppnp.2005.01.029,
arXiv:hep-ph/0501087.

\bibitem{Iancu:2001}
E.~Iancu, A.~Leonidov and L.~McLerran,
``The renormalization group equation for the color glass condensate,''
Phys.\ Lett.\ B \textbf{510} (2001) 133--144,
doi:10.1016/S0370-2693(01)00524-X,
arXiv:hep-ph/0102009.



\bibitem{Wiese:2016}
K.~J.~Wiese,
``Coherent-state path integral versus coarse-grained effective stochastic equation
of motion: From reaction diffusion to stochastic sandpiles,''
Phys.\ Rev.\ E \textbf{93} (2016) 042117,
doi:10.1103/PhysRevE.93.042117,
arXiv:1501.06514 [cond-mat.stat-mech].

\bibitem{Andreanov:2006}
A.~Andreanov, G.~Biroli, J.-P.~Bouchaud and A.~Lef\`evre,
``Field theories and exact stochastic equations for interacting particle systems,''
Phys.\ Rev.\ E \textbf{74} (2006) 030101(R),
doi:10.1103/PhysRevE.74.030101,
arXiv:cond-mat/0602307.



\bibitem{Elgart:2004}
V.~Elgart and A.~Kamenev,
``Rare event statistics in reaction-diffusion systems,''
Phys.\ Rev.\ E \textbf{70} (2004) 041106,
doi:10.1103/PhysRevE.70.041106,
arXiv:cond-mat/0404241.


\bibitem{Froissart:1961} M. Froissart. ``Asymptotic behavior and subtractions in the Mandelstam representation. ''\textit{Phys. Rev.}, 123:1053--1057, 1961.


\bibitem{Breakstone:1985}
A.~Breakstone \textit{et al.},
``Measurement of $\bar p p$ and $pp$ Elastic Scattering in the Dip Region at $\sqrt{s}=53$ GeV,''
Phys.\ Rev.\ Lett.\ \textbf{54} (1985) 2180--2183,
doi:10.1103/PhysRevLett.54.2180.


\bibitem{Block:2006}
M.~M.~Block,
``Hadronic forward scattering: Predictions for the Large Hadron Collider and cosmic rays,''
Phys.\ Rept.\ \textbf{436} (2006) 71--215,
doi:10.1016/j.physrep.2006.06.003.

\bibitem{Dosch:1994}
H.~G.~Dosch, E.~Ferreira and A.~Kr\"amer,
``Nonperturbative QCD treatment of high-energy hadron-hadron scattering,''
Phys.\ Rev.\ D \textbf{50} (1994) 1992--2015,
doi:10.1103/PhysRevD.50.1992.



\bibitem{Aad:2023}
G.~Aad \textit{et al.} [ATLAS Collaboration],
``Measurement of the total cross section and $\rho$-parameter from elastic
scattering in $pp$ collisions at $\sqrt{s}=13$ TeV with the ATLAS detector,''
Eur.\ Phys.\ J.\ C \textbf{83} (2023) 441,
doi:10.1140/epjc/s10052-023-11436-8.

\bibitem{Kohara:2017ats}
A.~K.~Kohara, E.~Ferreira,  T.~Kodama and M.~Rangel,
``Elastic amplitudes studied with the LHC measurements at 7 and 8 TeV,''
\textit{Eur.\ Phys.\ J.\ C} \textbf{77} (2017) 877,
doi:10.1140/epjc/s10052-017-5403-7.






\bibitem{Kohara:2019}
A.~K.~Kohara, E.~Ferreira and M.~Rangel,
``The interplay of hadronic amplitudes and Coulomb phase in LHC measurements at 13 TeV,''
Phys.\ Lett.\ B \textbf{789} (2019) 1--6,
doi:10.1016/j.physletb.2018.12.021,
arXiv:1811.03212 [hep-ph].

\bibitem{Kohara:2022ktp}
A. K. Kohara,
``Observation of two zeros of the real amplitude in $pp$ scattering at LHC energies,''
\textit{Eur.\ Phys.\ J.\ C} \textbf{83} (2023) 126,
doi:10.1140/epjc/s10052-023-11286-4.


\bibitem{Antchev:2019b}
G.~Antchev \textit{et al.} [TOTEM Collaboration],
``Elastic differential cross-section measurement at $\sqrt{s}=13$ TeV by TOTEM,''
Eur.\ Phys.\ J.\ C \textbf{79} (2019) 861,
doi:10.1140/epjc/s10052-019-7346-7,
arXiv:1812.08283 [hep-ex].



\bibitem{Cui:2023}
Z.-F.~Cui, D.~Binosi, C.~D.~Roberts, S.~M.~Schmidt and D.~N.~Triantafyllopoulos,
``Fresh look at experimental evidence for odderon exchange,''
Phys.\ Lett.\ B \textbf{839} (2023) 137826,
doi:10.1016/j.physletb.2023.137826,
arXiv:2205.15438 [hep-ph].



\bibitem{Munier:2003}
S.~Munier and R.~Peschanski,
``Geometric scaling as traveling waves,''
Phys.\ Rev.\ Lett.\ \textbf{91} (2003) 232001,
doi:10.1103/PhysRevLett.91.232001,
arXiv:hep-ph/0309177.

\bibitem{Kovner:2007}
A.~Kovner and M.~Lublinsky,
``Odderon and seven Pomerons: QCD Reggeon field theory from JIMWLK evolution,''
JHEP \textbf{02} (2007) 058,
doi:10.1088/1126-6708/2007/02/058,
arXiv:hep-ph/0512316.





\end{thebibliography}

\end{document}